\documentclass{aastex}%
\usepackage{amsmath}
\usepackage{amssymb}
\usepackage{amsfonts}
\usepackage{graphicx}%
\begin{document}
{\LARGE Influence of the back-reaction of streaming cosmic rays on magnetic
field generation and thermal instability }

\bigskip

\ \ \ \ \ \ \ \ \ \ \ \ \ \ \ \ \ \ \ \ \ \ {\large Anatoly K. Nekrasov}$^{1}
${\large \ and Mohsen Shadmehri}$^{2}$

$^{1}$ Institute of Physics of the Earth, Russian Academy of Sciences, 123995
Moscow, Russia;

anekrasov@ifz.ru, nekrasov.anatoly@gmail.com

$^{2}$ Department of Physics, Golestan University, Basij Square, Gorgan, Iran;

m.shadmehri@gu.ac.ir

\bigskip

\ \ \ \ \ \ \ \ \ \ \ \ \ \ \ \ \ \ \ \ \ \ \ \ \ \ \ \ \ \ \ \ \ \ \ \ \ \ 

\textbf{ABSTRACT}\ \ \ \ \ \ \ \ \ \ \ \ \ \ \ \ \ \ \ \ \ \ \ \ \ \ \ \ \ \ \ \ \ \ \ \ \ \ \ \ \ \ \ \ \ \ \ \ \ 

Using a multi-fluid approach, we investigate streaming and thermal
instabilities of the electron-ion plasma with homogeneous cold cosmic rays
propagating perpendicular to the background magnetic field. Perturbations are
considered to be also across the magnetic field. The back-reaction of cosmic
rays resulting in strong streaming instabilities is taken into account. It is
shown that for sufficiently short wavelength perturbations, the growth rates
can exceed the growth rate of cosmic-ray streaming instability along the
magnetic field found by Nekrasov \& Shadmehri (2012), which is in its turn
considerably larger than the growth rate of the Bell instability (2004). The
thermal instability is shown not to be subject to the action of cosmic rays in
the model under consideration. The dispersion relation for the thermal
instability has been derived which includes sound velocities of plasma and
cosmic rays, Alfv\'{e}n and cosmic-ray streaming velocities. The relation
between these parameters determines the kind of thermal instability ranging
from the Parker (1953) to the Field (1965) instabilities. The results obtained
can be useful for a more detailed investigation of electron-ion\textbf{\ }%
astrophysical objects\textbf{\ }such as supernova remnant shocks, galaxy
clusters and others including the dynamics of streaming cosmic rays.

\bigskip

\section{Introduction}

Cosmic rays are an important ingredient in astrophysical environments (see,
e.g., Zweibel 2003). They are capable of affecting the dynamics of
astrophysical plasma media leading to plasma heating, increasing the level of
ionization, driving outflows, modifying shocks, and so on (Zweibel 2003; Field
et al. 1969; Guo \& Oh 2008; Everett et al. 2008; Beresnyak et al. 2009; Samui
et al. 2010; En\ss lin et al. 2011). Cosmic-ray ionization contributes to star
formation (e.g., Yusef-Zadeh et al. 2007) and coupling of gas to the magnetic
field in accretion disks (Gammie 1996).

Thermal instability (Field 1965) have been used to explain existence of the
cold dense structures in the interstellar (Field 1965; Begelman \& McKee 1990;
Koyama \& Inutsuka 2000; Hennebelle \& P\'{e}rault 2000; S\'{a}nchez-Salcedo
et al. 2002; V\'{a}zquez-Semadeni et al. 2006; Fukue and Kamaya 2007; Inoue \&
Inutsuka 2008; Shadmehri et al. 2010) and intracluster (ICM; Field 1965;
Mathews \& Bregman 1978; Balbus \& Soker 1989; Loewenstein 1990;
Bogdanovi\'{c} et al. 2009; Parrish et al. 2009; Sharma et al. 2010) media.
For example, molecular filaments have been observed in galaxy clusters by
Conselice et al. (2001), Salom\'{e} et al. (2006), Cavagnolo et al. (2008),
and O'Dea et al. (2008).

In galaxy clusters, cosmic rays are wide spread (e.g., Guo \& Oh 2008;
En\ss lin et al. 2011). Therefore, they could exert influence on thermal
instability. In particular, including cosmic rays is required to explain the
atomic and molecular lines observed in filaments in clusters of galaxies by
Ferland et al. (2009). Such an investigation has been performed by Sharma et
al. (2010) in the framework of magnetohydrodynamic (MHD) equations. Numerical
analysis has shown that the cosmic-ray pressure can elongate cold filaments
along the magnetic field lines. However in general, cosmic rays can be
relativistic and have the streaming velocity of the order of the speed of
light and the mean energy larger than the particle rest energy. The
interaction of such particles with the thermal plasma can not be considered in
the framework of the conventional MHD.

It is well-known that the cosmic-ray drift current results in arising of the
return current in the background plasma and of streaming instabilities
generating magnetic fields (Achterberg 1983; Zweibel 2003; Bell 2004, 2005;
Riquelme \& Spitkovsky 2009, 2010). In papers by Achterberg (1983), Zweibel
(2003), and Bell (2004), the kinetic consideration of circularly-polarized
electromagnetic waves traveling along the background magnetic field where
cosmic rays also drift along the latter has been provided. For the case of the
large cosmic-ray Larmor radius in comparison with the wavelength, Bell (2004)
has found the growth rate some larger than that for the resonant cyclotron
instability proposed long time ago by Kulsrud and Pearce (1969). The general
case for the arbitrary mutual orientation of the background magnetic field,
the cosmic-ray current and the wave vector of perturbations has been
considered by Bell (2005) within the MHD framework. Riquelme \& Spitkovsky
(2010) have explored the case in which the cosmic-ray current is perpendicular
to the initial magnetic field and perturbations are excited along the latter.
In papers by Bell (2004, 2005) and Riquelmi \& Spitkovsky (2010),
instabilities were excited due to the return plasma current and obtained
growth rates were of the same order of magnitude. The dynamics of cosmic rays
did not play the role (in the analytical consideration). Nekrasov \& Shadmehri
(2012) have included the back-reaction of cosmic rays in the multi-fluid
approach for the model by Riquelmi \& Spitkovsky (2010) and found the growth
rate for the streaming instability considerably larger than that of Bell
(2004, 2005) and of Riquelmi \& Spitkovsky (2010) by a factor of the square
root from the ratio of plasma to cosmic-ray number densities. The second
result obtained by Nekrasov \& Shadmehri (2012) was that the thermal
instability is not subject to the action of cosmic rays in the model
considered. Instabilities along the background magnetic field driven by the
back-reaction of relativistic cosmic rays drifting also parallel to the
magnetic field have been considered by Nekrasov (2013).

These\textbf{\ }findings motivated us to investigate the case in which
perturbations arise transversely to the ambient magnetic field in the
directions both along and across the perpendicular cosmic-ray current. Such a
current can appear due to diamagnetic drift of cosmic rays and inhomogeneity
of the magnetic field (Bell 2005), due to gravitational cosmic-ray drift in
magnetic field and so on. Riquelmi \& Spitkovsky (2010) have discussed a
possibility of an appearance of the perpendicular cosmic-ray current because
of the magnetic wall effect of low-energy magnetized cosmic rays in the
pre-amplified \ magnetic fields in the upstream medium of supernova remnant
shocks. We note that such a mechanism can also operate in other cases in which
cosmic rays encounter magnetic clouds. As it follows from Bell (2005), where
the one-fluid MHD equations are used, the streaming instability does not exist
for perturbations perpendicular to the magnetic field. However, this result is
incorrect in the multi-fluid consideration (for three and more species) that
is shown in this paper and has been obtained earlier (e.g., Nekrasov 2007). We
here include the induced return current of the background plasma and
back-reaction of cosmic rays. In this approach, dispersion relations are
derived and growth rates are found analytically. We also consider possible
effects of cosmic rays on the thermal instability for the geometry under
consideration. We provide a comparison of results obtained in this paper with
those of Nekrasov \& Shadmehri (2012) and show a difference between them.

The paper is organized as follows. Section 2 contains the fundamental
equations for plasma, cosmic rays, and electromagnetic fields used in this
paper. The zero order state is discussed in Section 3. Wave equations are
given in Section 4. In Sections 5 and 6, the dispersion relations including
the plasma return current, cosmic-ray back-reaction and the terms describing
the thermal instability are derived and their solutions are found for
perturbations along and across the cosmic-ray current, respectively.
Discussion of important results obtained and possible astrophysical
implications are provided in Section 7. Conclusive remarks are summarized in
Section 8.

\bigskip

\section{Basic equations for a plasma and cosmic rays}

The fundamental equations for a plasma are the following:%
\begin{equation}
\frac{\partial\mathbf{v}_{j}}{\partial t}+\mathbf{v}_{j}\cdot\mathbf{\nabla
v}_{j}=-\frac{\mathbf{\nabla}p_{j}}{m_{j}n_{j}}+\frac{q_{j}}{m_{j}}%
\mathbf{E+}\frac{q_{j}}{m_{j}c}\mathbf{v}_{j}\times\mathbf{B},
\end{equation}
the equation of motion,%
\begin{equation}
\frac{\partial n_{j}}{\partial t}+\mathbf{\nabla}\cdot n_{j}\mathbf{v}_{j}=0,
\end{equation}
the continuity equation,%
\begin{equation}
\frac{\partial T_{i}}{\partial t}+\mathbf{v}_{i}\cdot\mathbf{\nabla}%
T_{i}+\left(  \gamma-1\right)  T_{i}\mathbf{\nabla}\cdot\mathbf{v}%
_{i}=-\left(  \gamma-1\right)  \frac{1}{n_{i}}%
\mathcal{L}%
_{i}\left(  n_{i},T_{i}\right)  +\nu_{ie}^{\varepsilon}\left(  n_{e}%
,T_{e}\right)  \left(  T_{e}-T_{i}\right)
\end{equation}
and%
\begin{equation}
\frac{\partial T_{e}}{\partial t}+\mathbf{v}_{e}\cdot\mathbf{\nabla}%
T_{e}+\left(  \gamma-1\right)  T_{e}\mathbf{\nabla}\cdot\mathbf{v}%
_{e}=-\left(  \gamma-1\right)  \frac{1}{n_{e}}%
\mathcal{L}%
_{e}\left(  n_{e},T_{e}\right)  -\nu_{ei}^{\varepsilon}\left(  n_{i}%
,T_{e}\right)  \left(  T_{e}-T_{i}\right)
\end{equation}
are the temperature equations for ions and electrons. In Equations (1) and
(2), the index $j=i,e$ denotes the ions and electrons, respectively. Notations
in Equations (1)-(4) are the following: $q_{j}$ and $m_{j}$ are the charge and
mass of species $j$; $\mathbf{v}_{j}$ is the hydrodynamic velocity; $n_{j}$ is
the number density; $p_{j}=n_{j}T_{j}$ is the thermal pressure; $T_{j}$ is the
temperature; $\nu_{ie}^{\varepsilon}(n_{e},T_{e})$ ($\nu_{ei}^{\varepsilon
}\left(  n_{i},T_{e}\right)  $) is the frequency of the thermal energy
exchange between ions (electrons) and electrons (ions) being $\nu
_{ie}^{\varepsilon}(n_{e},T_{e})=2\nu_{ie}$, where $\nu_{ie}$ is the collision
frequency of ions with electrons (Braginskii 1965); $n_{i}\nu_{ie}%
^{\varepsilon}\left(  n_{e},T_{e}\right)  =n_{e}\nu_{ei}^{\varepsilon}\left(
n_{i},T_{e}\right)  $; $\gamma$ is the ratio of the specific heats;
$\mathbf{E}$\textbf{\ }and $\mathbf{B}$ are the electric and magnetic fields;
and $c$ is the speed of light in vacuum. We include the thermal energy
exchange because the corresponding frequency $\nu_{ie}^{\varepsilon}$
($\nu_{ei}^{\varepsilon}$) must be compared with the dynamical frequency for
thermal instability. The cooling and heating of plasma species in Equations
(3) and (4) are described by the function $%
\mathcal{L}%
_{j}(n_{j},T_{j})=n_{j}^{2}\Lambda_{j}\left(  T_{j}\right)  -n_{j}\Gamma_{j}$,
where $\Lambda_{j}$ and $\Gamma_{j}$ are the cooling and heating functions,
respectively. This function has some deviation from the usually used
cooling-heating function $\pounds $ (Field 1965). Both functions are connected
to each other via the equality $%
\mathcal{L}%
_{j}\left(  n_{j},T_{j}\right)  =m_{j}n_{j}\pounds _{j}$. Our choice is
analogous to those of Begelman \& Zweibel (1994), Bogdanovi\'{c} et al.
(2009), and Parrish et al. (2009). The function $\Lambda_{j}\left(
T_{j}\right)  $ can be found, for example, in Tozzi \& Norman (2001). We do
not take into account the transverse thermal fluxes in the temperature
equations, which are small in a weekly collisional plasma (Braginskii 1965)
being considered in this paper. For simplicity, we do not take into account a
collisional coupling of ions and electrons in Equation (1). The corresponding
condition will be given in Section 7.

The cosmic rays that we are interested in here, are considered as a possible
source of the magnetic field generation and amplification in different
astrophysical environments in which cosmic-ray fluxes may exist (Zweibel \&
Everett 2010) as well as their possible influence on thermal instability. It
is important that cosmic rays have a drift velocity or a current relative to
the direction of the background magnetic field and can excite instabilities
due to their streaming. In this case, we are not interested in the cosmic-ray
history, i.e., in the spatial and momentum diffusion of the quasi-isotropic
cosmic-ray distribution function, described by the transport equation in the
turbulent medium (e.g., Skilling 1975), and consider cosmic rays as beams
governed by MHD equations in the vicinity of their local sources. Such an
approach is adopted in the beam-plasma systems to study streaming
instabilities.\textbf{\ }Equations for relativistic cosmic rays which can be
in general both protons and electrons, we apply in the form of relativistic
MHD equations given by Lontano et al. (2001)
\begin{equation}
\frac{\partial\left(  R_{cr}\mathbf{p}_{cr}\right)  }{\partial t}%
+\mathbf{v}_{cr}\cdot\mathbf{\nabla}\left(  R_{cr}\mathbf{p}_{cr}\right)
=-\frac{\mathbf{\nabla}p_{cr}}{n_{cr}}+q_{cr}\left(  \mathbf{E+}\frac{1}%
{c}\mathbf{v}_{cr}\times\mathbf{B}\right)  ,
\end{equation}%
\begin{equation}
\left(  \frac{\partial}{\partial t}+\mathbf{v}_{cr}\cdot\mathbf{\nabla
}\right)  \left(  \frac{p_{cr}\gamma_{cr}^{\Gamma_{cr}}}{n_{cr}^{\Gamma_{cr}}%
}\right)  =0,
\end{equation}
where%
\begin{equation}
R_{cr}=1+\frac{\Gamma_{cr}}{\Gamma_{cr}-1}\frac{T_{cr}}{m_{cr}c^{2}}.
\end{equation}
In these equations, $\mathbf{p}_{cr}=\gamma_{cr}m_{cr}\mathbf{v}_{cr}$ is the
momentum of a cosmic-ray particle having the rest mass $m_{cr}$ and velocity
$\mathbf{v}_{cr}$, $q_{cr}$ is its charge, $p_{cr}=\gamma_{cr}^{-1}%
n_{cr}T_{cr}$ is the kinetic pressure, $n_{cr}$ is the number density in the
laboratory frame, $\Gamma_{cr}$ is the adiabatic index, $\gamma_{cr}=\left(
1-\mathbf{v}_{cr}^{2}/c^{2}\right)  ^{-1/2}$ is the relativistic factor. The
continuity equation for cosmic rays is the same as Equation (2) at $j=cr$.
Equation (7) can be used for both cold nonrelativistic, $T_{cr}\ll$
$m_{cr}c^{2}$, and hot relativistic, $T_{cr}\gg$ $m_{cr}c^{2}$, cosmic rays.
In the first (second) case, we have $\Gamma_{cr}=5/3$ ($4/3$) (Lontano et al.
2001). The general form of the value $R_{cr}$, which is valid for any
relations between $T_{cr}$ and $m_{cr}c^{2}$, can be found, e.g., in Toepfer
(1971) and Dzhavakhishvili \& Tsintsadze (1973). We note that relativistic MHD
equations are obtained from the kinetic equations for species (e.g., Toepfer
1971; Dzhavakhishvili \& Tsintsadze 1973) and the form of Equations (5)-(7) is
equivalent to equations for cosmic rays used in other papers (e.g., Sakai \&
Kawata 1980; Mikhailovskii et al. 1985; Mofiz \& Khan 1993; Gratton et al.
1998; Haim 2009).\textbf{\ }It should be\textbf{\ }noted that in general the
notation $T_{cr}$ is considered not to be as the temperature, but as some
typical internal energy of the cosmic-ray distribution. To avoid confusion,
this notation could be changed via $p_{cr}$. However, we retain it as it is
given in (Lontano et al. 2001). The relativistic MHD\ equation (5) has a
general form and can be also applied to nonrelativistic and relativistic fluid
flows or beam particles (see, e.g., Toepfer 1971; Wallis et al. 1975;
Hazeltine \& Mahajan 2000; Haim 2009). We note that in the multifluid part of
their paper (Appendix A), Riquelme \& Spitkovsky (2010) have used the equation
for cosmic rays (Equation (A1)) analogous to Equation (5) in the cold
temperature regime with a beam velocity. We also note that the simple
one-fluid MHD equations have been used by Sharma et al. (2009, 2010) to
consider the influence of adiabatic cosmic rays with the diffusive energy flux
on the buoyancy and thermal instabilities in galaxy clusters, correspondingly.

Equations (1)-(6) are solved together with Maxwell's equations
\begin{equation}
\mathbf{\nabla\times E=-}\frac{1}{c}\frac{\partial\mathbf{B}}{\partial t}%
\end{equation}
and
\begin{equation}
\mathbf{\nabla\times B=}\frac{4\pi}{c}\mathbf{j+}\frac{1}{c}\frac
{\partial\mathbf{E}}{\partial t},
\end{equation}
where $\mathbf{j=j}_{pl}+\mathbf{j}_{cr}=\sum_{j}q_{j}n_{j}\mathbf{v}%
_{j}+\mathbf{j}_{cr}$. We note that Gauss' law for $\mathbf{B}$ is
automatically followed from Equation (8), and Gauss' law for $\mathbf{E}$ is
automatically obtained from Equations (2) and (9).  \textbf{\ }

\bigskip

\section{Zero order system state}

It is known for a long time that a return current is induced in a plasma
penetrated by an external beam current (Roberts \& Bennett 1968). The return
plasma current equal to the external one and directed oppositely arises due to
self-consistent electromagnetic perturbations of plasma under the action of an
external current (e.g., Cox \& Bennett 1970; Hammer \& Rostoker 1970; Berk \&
Pearlstein 1976). As a result, the condition of quasineutrality and the
absence of the total current are maintained. In astrophysical plasmas, such
external beam currents are cosmic-ray flows. In papers devoted to cosmic-ray
streaming instabilities in the situation where drift velocities of plasma
species and cosmic rays are directed along the background magnetic field
(e.g., Achterberg 1983; Zweibel 2003; Bell 2004; Riquelmi \& Spitkovsky 2009;
Nekrasov 2013), it has been also assumed that to zeroth-order the system is of
charge neutrality and there is no net current due to appearance of the plasma
return current. Here, we consider other situation in which cosmic rays can
drift across the background magnetic field. One such a possibility has been
considered by Riquelme \& Spitkovsky (2010) for the upstream medium of
supernova remnant shocks. It was shown that near the shock cosmic rays having
the Larmor radius smaller than the length scale of pre-amplified,
quasi-transverse magnetic field generated by the highest energy cosmic rays
due to the Bell instability (Bell 2004), will produce\textbf{ }a current
perpendicular to the initial, pre-amplified field due to the coherent
deflection in the \textquotedblleft homogeneous\textquotedblright\ (large
scale) magnetic field (see Riquelme \& Spitkovsky 2010 for details). One can
say that this perpendicular current arises due to the magnetic wall effect.
Therefore, we would like to note that such a mechanism could also occur in
other astrophysical environments where cosmic rays can encounter magnetic
fields (clouds). Two-dimensional particle-in-cell simulations (Riquelme \&
Spitkovsky 2010) have confirmed the formation of the perpendicular mean
cosmic-ray velocity ( at $\sim c/2$).

As in the case of cosmic rays drifting along the magnetic field, one can also
assume the generation of the return plasma current compensating the
perpendicular cosmic-ray one. It can be shown that in the ideal model of
Riquelme \&\ Spitkovsky (2010), we have an infinite sheet cosmic-ray current,
which forms a homogeneous magnetic field parallel to the current plane and
perpendicular to the current direction. In this case, the return current can
be only produced by the time-dependent perpendicular electric field in the
zero order state, in which plasma species experience a polarization drift
across the magnetic field. 

Let us find this electric field. We consider a uniform plasma embedded in the
uniform magnetic field $\mathbf{B}_{0}$,\textbf{\  }$\mathbf{j}_{cr0}%
$\textbf{\ }is directed along the $y$-axis. From Equations (1) and (5), where
we take into account the electric and polarization drifts of particles, and
from Equation (9) without left-hand side and with account for the displacement
current, one can find the time-dependent zero order electric field\textbf{
}$\mathbf{E}_{0}$\textbf{\ }defined by\textbf{ }%
\begin{equation}
\frac{\partial\mathbf{E}_{0}}{\partial t}=-4\pi\mathbf{j}_{cr0}\frac
{c_{Ai}^{2}}{c_{Ai}^{2}+c^{2}},
\end{equation}
where $c_{Ai}=\left(  B_{0}^{2}/4\pi m_{i}n_{i0}\right)  ^{1/2}$\ is the ion
Alfv\'{e}n velocity. The conditions $\partial/\partial t\ll\omega_{cj}$\ and
$R_{cr}\gamma_{cr}\partial/\partial t\ll\omega_{ccr}$, where $\omega_{cj}%
$\ $=q_{j}B_{0}/m_{j}c$\ is the cyclotron frequency, and the condition of
quasi-neutrality, $q_{i}n_{i0}+q_{e}n_{e0}+q_{cr}n_{cr0}=0$\ (the number
density $n_{cr}$\ is the one in the laboratory frame), have been used. The
polarization drift of cosmic rays in Equation (10) has been omitted. This
equation in the case $c^{2}\gg c_{Ai}^{2}$\ has been given by Riquelme \&
Spitkovsky (2010) without derivation. We note that at the absence of the
background plasma ($n_{i0}\rightarrow0,$\ $c_{Ai}\rightarrow\infty
$\ ),\textbf{ }Equation (10) results in Maxwell's equation\textbf{ }%
$4\pi\mathbf{j}_{cr0}+\partial\mathbf{E}_{0}/\partial t=0$\textbf{\ }for the
uniform magnetic field. Using Equation (10) in the limit $c^{2}\gg c_{Ai}^{2}%
$, we find the return plasma current\textbf{ }$\mathbf{j}_{ret}$ defined by
the polarization drift of ions\textbf{ }$\mathbf{u}_{pl}$
\begin{equation}
\mathbf{j}_{ret}=q_{i}n_{i0}\mathbf{u}_{pl}=-\mathbf{j}_{cr0},
\end{equation}
whose magnitude is equal to the cosmic-ray current and has the opposite
direction. The polarization drift of electrons is not taken into account
because of a\ small electron mass. In general, the zero order electric
field\textbf{ }$\mathbf{E}_{0}$\textbf{\ }cannot operates indefinitely. This
field continues only during the action of cosmic rays. If we put, for
convenience, $\mathbf{j}_{cr0}=$ $q_{cr}n_{cr0}\mathbf{u}_{cr}$, where
$\mathbf{u}_{cr}$ is the velocity of cosmic rays along the $y$-axis, then we
obtain from Equation (11) that $\mathbf{u}_{pl}=-\left(  q_{cr}n_{cr0}%
/q_{i}n_{i0}\right)  \mathbf{u}_{cr}$. Thus, $u_{pl}\ll u_{cr}$ because
$n_{cr0}\ll n_{i0}$. Below, the plasma drift velocity $\mathbf{u}_{pl}$ will
be also taken into account together with $\mathbf{u}_{cr}$.

Above in this Section, we have discussed a zero order state for a model
considered by Riquelme \& Spitkovsky (2010) in which cosmic-ray and plasma
return currents are perpendicular to the background magnetic field. However,
perpendicular currents can also form due to other reasons. For example, cosmic
rays and plasma charged species can drift across the magnetic field, which is
inhomogeneous in the longitudinal and/or transverse directions, and in the
presence of a perpendicular gravitational acceleration. In this case, we
think, a return current cannot appear because cosmic rays are not an external
agent penetrating a plasma. Further, the large energy cosmic rays having the
Larmor radius much larger than inhomogeneities of magnetic force lines can
result in a transverse current. It is possible that in this case the return
current can arise. Also, diamagnetic drifts due to transverse pressure
gradients produce transverse currents. 

For simplicity, we further consider the case in which background temperatures
of electrons and ions are equal each other, i.e. $T_{e0}=T_{i0}=T_{0}$. The
case $T_{e0}\neq T_{i0}$\ for thermal instability has been considered, for
instance, by Nekrasov (2011, 2012). Here, we will omit the perturbed terms
$\propto$\ $\left(  T_{e0}-T_{i0}\right)  $ in the temperature equations.
However, to follow the symmetric contribution of ions and electrons in a
convenient way, we make some calculations by assuming different temperatures.
Then, thermal equations (3) and (4) in the background state take the form%

\begin{equation}%
\mathcal{L}%
_{i}\left(  n_{i0},T_{i0}\right)  =%
\mathcal{L}%
_{e}\left(  n_{e0},T_{e0}\right)  =0.
\end{equation}

\bigskip

\section{Wave\ equations}

For perturbations across the background magnetic field when $\partial/\partial
z=0$, Equations (8) and (9) give us the following two equations:%
\begin{equation}
c^{2}\left(  \frac{\partial}{\partial t}\right)  ^{-2}\left(  \frac
{\partial^{2}E_{1x}}{\partial y^{2}}-\frac{\partial^{2}E_{1y}}{\partial
x\partial y}\right)  -E_{1x}=4\pi\left(  \frac{\partial}{\partial t}\right)
^{-1}j_{1x}%
\end{equation}
and
\begin{equation}
c^{2}\left(  \frac{\partial}{\partial t}\right)  ^{-2}\left(  -\frac
{\partial^{2}E_{1x}}{\partial x\partial y}+\frac{\partial^{2}E_{1y}}{\partial
x^{2}}\right)  -E_{1y}=4\pi\left(  \frac{\partial}{\partial t}\right)
^{-1}j_{1y},
\end{equation}
where $\mathbf{j}_{1}=\mathbf{j}_{pl1}+\mathbf{j}_{cr1}$ and the subscript $1$
here and below denotes the perturbed values. The third equation describes the
ordinary electromagnetic wave with $\mathbf{E}_{1}\mathbf{\parallel B}_{0}$
and is split from Equations (13) and (14). The general expressions for the
components $j_{pl1x,y}$ and $j_{cr1x,y}$ are given in the Appendices A and B
(Equations (A54)-(A56) and (B19)-(B21)). These expressions are available for
both magnetized and non-magnetized systems, electron-positron, pair-ion and
dusty plasmas and so on. Besides, they include the radiation-condensation
effects. In their general form, these expressions are very complicated.
Therefore to proceed analytically, one must apply simplifying assumptions. We
are interested in magnetized systems consisting of electrons, ions, and cosmic
rays, in which cyclotron frequencies of species are much larger than the
Doppler-shifted dynamical frequencies. In our case, this implies
\begin{align}
\omega_{ci}^{2} &  \gg\left(  \frac{\partial}{\partial t}+u_{pl}\frac
{\partial}{\partial y}\right)  ^{2},\\
\omega_{ccr}^{2} &  \gg\gamma_{cr0}^{4}\left(  \frac{\partial}{\partial
t}+u_{cr}\frac{\partial}{\partial y}\right)  ^{2}\nonumber
\end{align}
(see Equations (A5), (A8), and (B7)). As we have noted above, cosmic rays can
be both protons and electrons. For ultrarelativistic cosmic rays,
$\gamma_{cr0}\gg1$, the second Equation (15) can be violated. Such a
case\textbf{\ }in which cosmic rays become unmagnetized is not considered in
this paper. We here assume also that the case $T_{cr}\ll m_{cr}c^{2}$ is
satisfied, i.e. cosmic rays are cold. Another condition that simplifies the
treatment considerably is to assume the wavelength of perturbations to be much
larger than the thermal Larmor radius of particles $\rho_{j}$%
\begin{align}
1 &  \gg\rho_{i}^{2}\mathbf{\nabla}^{2},\\
1 &  \gg\rho_{cr}^{2}\gamma_{cr0}\left(  \gamma_{cr0}^{2}\frac{\partial^{2}%
}{\partial x^{2}}+\frac{\partial^{2}}{\partial y^{2}}\right)  ,\nonumber
\end{align}
where $\rho_{i}\sim\left(  T_{i0}/m_{i}\omega_{ci}^{2}\right)  ^{1/2}$ and
$\rho_{cr}=c_{scr}/\omega_{ccr}$ (see Equations (A41) and (B11)). The
additional conditions for cosmic rays simplifying their contribution to a
current will be given below. The third simplification is to consider
perturbations along and across the cosmic-ray velocity $\mathbf{u}_{cr}$
separately. The first case is simpler. Therefore, we begin with its consideration.

\bigskip

\section{The case $\frac{\partial}{\partial y}\neq0,\frac{\partial}{\partial
x}=0$}

Using Equation (A56) and performing calculations of the corresponding
quantities, we find the components of the plasma dielectric permeability
tensor ($v_{i0y}$ has been changed by $u_{pl}$)%

\begin{align}
\varepsilon_{plxx}  &  =\frac{\omega_{pi}^{2}}{\omega_{ci}^{2}}\left(
\frac{\partial}{\partial t}+u_{pl}\frac{\partial}{\partial y}\right)
^{2}\left(  \frac{\partial}{\partial t}\right)  ^{-2}\\
&  -\frac{\omega_{pi}^{2}}{\omega_{ci}^{2}}\frac{1}{m_{i}}\left[
T_{i0}+T_{e0}-\frac{G_{1}+G_{3}}{D}\frac{\partial}{\partial t}-\frac
{G_{2}+G_{4}}{D}\left(  \frac{\partial}{\partial t}+u_{pl}\frac{\partial
}{\partial y}\right)  \right]  \frac{\partial^{2}}{\partial y^{2}}\left(
\frac{\partial}{\partial t}\right)  ^{-2},\nonumber\\
\varepsilon_{plxy}  &  =\left(  \frac{\omega_{pi}^{2}\omega_{ci}}{\Omega
_{i}^{2}}+\frac{\omega_{pe}^{2}\omega_{ce}}{\Omega_{e}^{2}}\right)  \left(
\frac{\partial}{\partial t}\right)  ^{-1}\nonumber\\
&  +\frac{\omega_{pi}^{2}}{\omega_{ci}^{3}}\frac{1}{m_{i}}\left[  T_{i0}%
-\frac{G_{2}+G_{4}}{D}\left(  \frac{\partial}{\partial t}+u_{pl}\frac
{\partial}{\partial y}\right)  \right]  \frac{\partial^{2}}{\partial y^{2}%
}\left(  \frac{\partial}{\partial t}\right)  ^{-1},\nonumber\\
\varepsilon_{plyx}  &  =-\left(  \frac{\omega_{pi}^{2}\omega_{ci}}{\Omega
_{i}^{2}}+\frac{\omega_{pe}^{2}\omega_{ce}}{\Omega_{e}^{2}}\right)  \left(
\frac{\partial}{\partial t}\right)  ^{-1}\nonumber\\
&  -\frac{\omega_{pi}^{2}}{\omega_{ci}^{3}}\frac{1}{m_{i}}\left[  T_{i0}%
-\frac{G_{3}}{D}\frac{\partial}{\partial t}-\frac{G_{4}}{D}\left(
\frac{\partial}{\partial t}+u_{pl}\frac{\partial}{\partial y}\right)  \right]
\frac{\partial^{2}}{\partial y^{2}}\left(  \frac{\partial}{\partial t}\right)
^{-1},\nonumber\\
\varepsilon_{plyy}  &  =\frac{\omega_{pi}^{2}}{\omega_{ci}^{2}}.\nonumber
\end{align}
For obtaining Equation (17), we have taken into account that $m_{i}\gg m_{e}$
and $n_{i0}\simeq n_{e0}$. Analogously from Equation (B21), we obtain the
cosmic-ray dielectric permeability tensor%

\begin{align}
\varepsilon_{crxx} &  =\frac{\omega_{pcr}^{2}}{\omega_{ccr}^{2}}\gamma
_{cr0}^{3}\left(  \frac{\partial}{\partial t}+u_{cr}\frac{\partial}{\partial
y}\right)  ^{2}\left(  \frac{\partial}{\partial t}\right)  ^{-2}-\frac
{\omega_{pcr}^{2}}{\omega_{ccr}^{2}}\gamma_{cr0}^{2}c_{scr}^{2}\ \left(
\frac{u_{cr}}{c^{2}}\frac{\partial}{\partial t}+\frac{\partial}{\partial
y}\right)  \frac{\partial}{\partial y}\left(  \frac{\partial}{\partial
t}\right)  ^{-2},\\
\varepsilon_{crxy} &  =-\varepsilon_{cryx}=\frac{\omega_{pcr}^{2}}{\Omega
_{cr}^{2}}\omega_{ccr}\left(  \frac{\partial}{\partial t}\right)  ^{-1}%
+\frac{\omega_{pcr}^{2}}{\omega_{ccr}^{3}}\gamma_{cr0}^{3}c_{scr}^{2}\ \left(
\frac{u_{cr}}{c^{2}}\frac{\partial}{\partial t}+\frac{\partial}{\partial
y}\right)  \frac{\partial}{\partial y}\left(  \frac{\partial}{\partial
t}\right)  ^{-1},\nonumber\\
\varepsilon_{cryy} &  =\frac{\omega_{pcr}^{2}}{\omega_{ccr}^{2}}\gamma
_{cr0}.\nonumber
\end{align}
Here, we have used the additional condition for cosmic rays%
\[
1\gg\gamma_{cr0}^{3}\rho_{cr}^{2}\frac{u_{cr}}{c^{2}}\left(  \frac{\partial
}{\partial t}+u_{cr}\frac{\partial}{\partial y}\right)  \frac{\partial
}{\partial y}%
\]
(see Equation (B11)). The term proportional to $u_{cr}/c^{2}$\ in Equation
(18) shows the contribution of $\gamma_{cr1}$\ to the cosmic-ray pressure
perturbation (see Equations (B8) and (B9)).

\bigskip

\subsection{Wave equation}

From Equations (13) and (14), using Equations (A54), (A55), (B19), and (B20)
and omitting the contribution of the displacement current under
conditon\textbf{ }$\varepsilon_{xx}\gg1$, we obtain equation
\begin{equation}
\varepsilon_{yy}c^{2}\left(  \frac{\partial}{\partial t}\right)  ^{-2}%
\frac{\partial^{2}E_{1x}}{\partial y^{2}}=\left(  \varepsilon_{xx}%
\varepsilon_{yy}-\varepsilon_{xy}\varepsilon_{yx}\right)  E_{1x},
\end{equation}
where $\varepsilon_{ij}=\varepsilon_{plij}+\varepsilon_{crij}$. The values
$\varepsilon_{ij}$ are defined by Equations (17) and (18). When calculating
the right-hand side of Equation (19), we assume some additional conditions
except those given by Equations (15) and (16). We will neglect the
contribution to $\varepsilon_{xy}\varepsilon_{yx}$ of the thermal cosmic-ray
term in $\varepsilon_{crxy}$ and $\varepsilon_{cryx}$. Besides, we will use
the condition of quasineutrality in $\varepsilon_{xy}$ and $\varepsilon_{yx}$
and neglect the terms arising due to expansion of $\Omega_{i,cr}^{-2}$. An
analysis shows that the corresponding conditions can be written in the form
\begin{align}
&  \min\left\{  \gamma_{cr0}\left(  \frac{\partial}{\partial t}+u_{cr}%
\frac{\partial}{\partial y}\right)  ^{2};c_{scr}^{2}\ \left(  \frac{u_{cr}%
}{c^{2}}\frac{\partial}{\partial t}+\frac{\partial}{\partial y}\right)
\frac{\partial}{\partial y}\right\}  \\
&  \gg\gamma_{cr0}^{3}\frac{c_{scr}^{4}}{\omega_{ccr}^{2}}\ \left(
\frac{u_{cr}}{c^{2}}\frac{\partial}{\partial t}+\frac{\partial}{\partial
y}\right)  ^{2}\frac{\partial^{2}}{\partial y^{2}};\gamma_{cr0}\frac
{c_{spl}^{2}c_{scr}^{2}}{\omega_{ci}\omega_{ccr}}\ \left(  \frac{u_{cr}}%
{c^{2}}\frac{\partial}{\partial t}+\frac{\partial}{\partial y}\right)
\frac{\partial^{3}}{\partial y^{3}};\nonumber\\
&  \gamma_{cr0}\frac{c_{scr}^{2}}{\omega_{ci}\omega_{ccr}}\ \left(
\frac{u_{cr}}{c^{2}}\frac{\partial}{\partial t}+\frac{\partial}{\partial
y}\right)  \frac{\partial}{\partial y}\left(  \frac{\partial}{\partial
t}+u_{pl}\frac{\partial}{\partial y}\right)  ^{2},\nonumber
\end{align}
where $c_{spl}=\left(  2\gamma T_{i0}/m_{i}\right)  ^{1/2}$. For simplicity,
for writing these inequalities, we considered the terms, in which plasma
frequencies $\omega_{pi}$ and $\omega_{pcr}$ are cancelled. In the term
$\varepsilon_{xx}$, we used a cosmic-ray term in the main. According to
Equations (16) and (20), the contribution of the term $\varepsilon
_{xy}\varepsilon_{yx}$ to Equation (19) is small. For example, an estimation
shows (without thermal terms in Equation (18)) that $\varepsilon
_{plxy}\varepsilon_{plyx}/\varepsilon_{plxx}\varepsilon_{plyy}\sim$\ $\left(
\partial/\partial t+u_{pl}\partial/\partial y\right)  ^{2}/\omega_{ci}^{2}%
\ll1$ and $\varepsilon_{crxy}\varepsilon_{cryx}/\varepsilon_{crxx}%
\varepsilon_{cryy}\sim$\ $\left(  \partial/\partial t+u_{cr}\partial/\partial
y\right)  ^{2}/\gamma_{cr}^{4}\omega_{ccr}^{2}\ll1$ (see Equation (15)). Thus,
we obtain the simple wave equation
\begin{equation}
c^{2}\frac{\partial^{2}E_{1x}}{\partial y^{2}}=\varepsilon_{xx}\left(
\frac{\partial}{\partial t}\right)  ^{2}E_{1x}.
\end{equation}
We note that for these perturbations\textbf{ }$E_{1y}=0$ and $B_{1x,y}=0$,
$B_{1z}\neq0$.

\bigskip

\subsection{Dispersion relation}

Using Equations (17) and (18) to find $\varepsilon_{xx}$ and accomplishing the
Fourier transform in Equation (21), we find for perturbations of the form
$\exp\left(  ik_{y}y-i\omega t\right)  $ the following dispersion relation:
\begin{align}
0  &  =\frac{\omega_{pi}^{2}}{\omega_{ci}^{2}}\left(  \omega-k_{y}%
u_{pl}\right)  ^{2}+\frac{\omega_{pcr}^{2}}{\omega_{ccr}^{2}}\gamma_{cr0}%
^{3}\left(  \omega-k_{y}u_{cr}\right)  ^{2}\\
&  -\frac{\omega_{pi}^{2}}{\omega_{ci}^{2}}k_{y}^{2}\frac{1}{m_{i}}\left[
T_{i0}+T_{e0}+\frac{G_{1}+G_{3}}{D}i\omega+\frac{G_{2}+G_{4}}{D}i\left(
\omega-k_{y}u_{pl}\right)  \right] \nonumber\\
&  -\frac{\omega_{pcr}^{2}}{\omega_{ccr}^{2}}\gamma_{cr0}^{2}k_{y}^{2}%
c_{scr}^{2}\ -k_{y}^{2}c^{2}.\nonumber
\end{align}
Below, we consider solutions of Equation (22) for the streaming instability
and an influence of the streaming and thermal pressure effects on the thermal instability.

\bigskip

\subsubsection{\textit{Streaming instability}}

Let us set in Equation (22) all frequencies $\Omega$ equal to zero. To be more
specific, it means that $\omega-k_{y}u_{pl}\gg\Omega_{T,ni},\Omega_{\epsilon}$
and $\omega\gg\Omega_{T,ne},\Omega_{\epsilon}$, where $\Omega_{ie}\simeq
\Omega_{ei}=\Omega_{\epsilon}$ (the frequencies $\Omega$ are defined by
Equation (A12)). These conditions mean that we consider perturbations much
faster than the typical time scales of thermal instability.\textbf{ }Then,
this equation takes the form%
\begin{align}
0 &  =\frac{\omega_{pi}^{2}}{\omega_{ci}^{2}}\left(  \omega-k_{y}%
u_{pl}\right)  ^{2}+\frac{\omega_{pcr}^{2}}{\omega_{ccr}^{2}}\gamma_{cr0}%
^{3}\left(  \omega-k_{y}u_{cr}\right)  ^{2}\\
&  -\left(  \frac{\omega_{pi}^{2}}{\omega_{ci}^{2}}c_{spl}^{2}+\frac
{\omega_{pcr}^{2}}{\omega_{ccr}^{2}}\gamma_{cr0}^{2}c_{scr}^{2}+c^{2}\right)
k_{y}^{2}.\nonumber
\end{align}
The solution of Equation (23) is the following:
\begin{equation}
\omega=\frac{k_{y}\left(  u_{pl}+du_{cr}\right)  }{1+d}\pm\frac{k_{y}}%
{1+d}\left[  -\left(  u_{cr}-u_{pl}\right)  ^{2}d+\left(  1+d\right)  \left(
c_{spl}^{2}+\gamma_{cr0}^{-1}dc_{scr}^{2}+c_{Ai}^{2}\right)  \right]  ^{1/2},
\end{equation}
where
\begin{equation}
d=\frac{\omega_{ci}^{2}}{\omega_{pi}^{2}}\frac{\omega_{pcr}^{2}}{\omega
_{ccr}^{2}}\gamma_{cr0}^{3}=\frac{m_{cr}}{m_{i}}\frac{n_{cr0}}{n_{i0}}%
\gamma_{cr0}^{3}.
\end{equation}
We see that the streaming instability has a threshold $u_{crth}$ defined by
the sound and ion Alfv\'{e}n velocities%
\begin{equation}
u_{crth}^{2}=\left(  1+d^{-1}\right)  \left(  c_{spl}^{2}+\gamma_{cr0}%
^{-1}dc_{scr}^{2}+c_{Ai}^{2}\right)  .
\end{equation}
When this threshold is exceeded, $u_{cr}^{2}\gg u_{crth}^{2}$, the growth rate
$\delta_{gr}$ is given by%
\begin{equation}
\delta_{gr}=\frac{d^{1/2}}{1+d}k_{y}u_{cr}.
\end{equation}
These perturbations move with the phase velocity $v_{ph}=\left(
u_{pl}+du_{cr}\right)  /\left(  1+d\right)  $. We see that the induced plasma
drift velocity $u_{pl}$ does not affect on the growth rate because $u_{pl}\ll
u_{cr}$ (see Equation (11)), but can contribute to the real part of the frequency.

\bigskip

\subsubsection{\textit{Thermal instability}}

We now take into account the terms describing the thermal instability in
Equation (22). We consider the fast thermal energy exchange regime in which
$\Omega_{\epsilon}\gg\partial/\partial t,\Omega_{Ti,e}$. Using Equations (A29)
and (A30), we have%
\begin{equation}
\frac{\gamma\left(  2\omega-k_{y}u_{pl}\right)  +i\Omega_{T,n}}{\gamma\left(
2\omega-k_{y}u_{pl}\right)  +i\gamma\Omega_{T}}=c_{spl}^{-2}\left(
du_{cr}^{2}-\gamma_{cr0}^{-1}dc_{scr}^{2}-c_{Ai}^{2}+\frac{\omega^{2}}%
{k_{y}^{2}}\right)  ,
\end{equation}
where%
\begin{align*}
\Omega_{T,n} &  =\Omega_{Te}+\Omega_{Ti}-\Omega_{ne}-\Omega_{ni},\\
\Omega_{T} &  =\Omega_{Te}+\Omega_{Ti}.
\end{align*}
When obtaining Equation (28), we have assumed $\omega\ll k_{y}u_{cr}$ that
physically corresponds to the low frequency thermal instability in a
comparison roughly with the streaming instability. If the right-hand side of
Equation (28) is much less than unity, we obtain Field's isobaric solution
$2\omega=k_{y}u_{pl}-i\Omega_{T,n}/\gamma$ (Field 1965). These perturbations
travel with the phase velocity $u_{pl}/2$. In the opposite case, Equation (28)
has the Parker's isochoric solution $2\omega=k_{y}u_{pl}-i\Omega_{T}$ (Parker
1953). Thus, the presence of streaming cosmic rays can only change the kind of
thermal instability, but not influence on its growth rates. When the
right-hand side of Equation (28) is of the order of unity, the limiting
solutions intermix.

\bigskip

\section{The case $\frac{\partial}{\partial x}\neq0,\frac{\partial}{\partial
y}=0$}

Calculating the components of the plasma dielectric permeability tensor given
by Equation (A56), we obtain%
\begin{align}
\varepsilon_{plxx} &  =\frac{\omega_{pi}^{2}}{\omega_{ci}^{2}},\\
\varepsilon_{plxy} &  =\frac{\omega_{pi}^{2}\omega_{ci}}{\Omega_{i}^{2}%
}\left(  \frac{\partial}{\partial t}\right)  ^{-1}+\frac{\omega_{pe}^{2}%
\omega_{ce}}{\Omega_{e}^{2}}\left(  \frac{\partial}{\partial t}\right)
^{-1}\nonumber\\
&  +\frac{\omega_{pi}^{2}}{\omega_{ci}^{3}}\left[  \frac{1}{m_{i}}\left(
T_{i0}-\frac{G_{3}+G_{4}}{D}\frac{\partial}{\partial t}\right)  \frac
{\partial}{\partial x}-\omega_{ci}u_{pl}\right]  \frac{\partial}{\partial
x}\left(  \frac{\partial}{\partial t}\right)  ^{-1},\nonumber\\
\varepsilon_{plyx} &  =-\frac{\omega_{pi}^{2}\omega_{ci}}{\Omega_{i}^{2}%
}\left(  \frac{\partial}{\partial t}\right)  ^{-1}-\frac{\omega_{pe}^{2}%
\omega_{ce}}{\Omega_{e}^{2}}\left(  \frac{\partial}{\partial t}\right)
^{-1}\nonumber\\
&  -\frac{\omega_{pi}^{2}}{\omega_{ci}^{3}}\left[  \frac{1}{m_{i}}\left(
T_{i0}-\frac{G_{2}+G_{4}}{D}\frac{\partial}{\partial t}\right)  \frac
{\partial}{\partial x}+\omega_{ci}u_{pl}\right]  \frac{\partial}{\partial
x}\left(  \frac{\partial}{\partial t}\right)  ^{-1},\nonumber\\
\varepsilon_{plyy} &  =\frac{\omega_{pi}^{2}}{\omega_{ci}^{2}}-\frac
{\omega_{pi}^{2}}{\omega_{ci}^{2}}\frac{1}{m_{i}}\left(  T_{i0}+T_{e0}%
-\frac{G_{1}+G_{2}+G_{3}+G_{4}}{D}\frac{\partial}{\partial t}\right)
\frac{\partial^{2}}{\partial x^{2}}\left(  \frac{\partial}{\partial t}\right)
^{-2}\nonumber\\
&  +\frac{\omega_{pi}^{2}}{\omega_{ci}^{2}}u_{pl}^{2}\frac{\partial^{2}%
}{\partial x^{2}}\left(  \frac{\partial}{\partial t}\right)  ^{-2}%
-\frac{\omega_{pi}^{2}}{\omega_{ci}^{3}}u_{pl}\frac{1}{m_{i}}\frac{G_{2}%
-G_{3}}{D}\frac{\partial^{3}}{\partial x^{3}}\left(  \frac{\partial}{\partial
t}\right)  ^{-1}.\nonumber
\end{align}
From Equation (B21) for cosmic rays, we have
\begin{align}
\varepsilon_{crxx} &  =\frac{\omega_{pcr}^{2}}{\omega_{ccr}^{2}}\gamma
_{cr0}^{3},\\
\varepsilon_{crxy} &  =\frac{\omega_{pcr}^{2}\omega_{ccr}}{\Omega_{cr}^{2}%
}\left(  \frac{\partial}{\partial t}\right)  ^{-1}+\frac{\omega_{pcr}^{2}%
}{\omega_{ccr}^{3}}\gamma_{cr0}^{3}\left(  \gamma_{cr0}^{2}c_{scr}^{2}%
\frac{\partial}{\partial x}-\omega_{ccr}u_{cr}\right)  \frac{\partial
}{\partial x}\left(  \frac{\partial}{\partial t}\right)  ^{-1},\nonumber\\
\varepsilon_{cryx} &  =-\frac{\omega_{pcr}^{2}\omega_{ccr}}{\Omega_{cr}^{2}%
}\left(  \frac{\partial}{\partial t}\right)  ^{-1}-\frac{\omega_{pcr}^{2}%
}{\omega_{ccr}^{3}}\gamma_{cr0}^{3}\left(  c_{scr}^{2}\frac{\partial}{\partial
x}+\omega_{ccr}u_{cr}\right)  \frac{\partial}{\partial x}\left(
\frac{\partial}{\partial t}\right)  ^{-1},\nonumber\\
\varepsilon_{cryy} &  =\frac{\omega_{pcr}^{2}}{\omega_{ccr}^{2}}\gamma
_{cr0}\left[  1+\gamma_{cr0}^{2}u_{cr}^{2}\frac{\partial^{2}}{\partial x^{2}%
}\left(  \frac{\partial}{\partial t}\right)  ^{-2}\right]  -\frac{\omega
_{pcr}^{2}}{\omega_{ccr}^{2}}c_{scr}^{2}\gamma_{cr0}^{2}\frac{\partial^{2}%
}{\partial x^{2}}\left(  \frac{\partial}{\partial t}\right)  ^{-2}.\nonumber
\end{align}
In this geometry, the additional simplifying condition for the cosmic-ray
contribution except of Equation (16) follows from Equation (B11)
\[
1\gg\gamma_{cr0}^{2}\frac{u_{cr}c_{scr}}{c^{2}}\rho_{cr}\frac{\partial
}{\partial x}.
\]
We note that the terms $\varepsilon_{plxy}(\varepsilon_{plyx})$ and
$\varepsilon_{crxy}(\varepsilon_{cryx})$ contain here large terms $\omega
_{ci}u_{pl}$ and $\omega_{ccr}u_{cr}$, respectively. 

\bigskip

\subsection{Wave equation}

In the case under consideration, the wave equation has the form%
\begin{equation}
\varepsilon_{xx}c^{2}\left(  \frac{\partial}{\partial t}\right)  ^{-2}%
\frac{\partial^{2}E_{1y}}{\partial x^{2}}=\left(  \varepsilon_{xx}%
\varepsilon_{yy}-\varepsilon_{xy}\varepsilon_{yx}\right)  E_{1y}.
\end{equation}
Using Equations (29) and (30) and calculating the right-hand side of Equation
(31), we find the simple expression for $\varepsilon_{xx}\varepsilon
_{yy}-\varepsilon_{xy}\varepsilon_{yx}$
\begin{align}
\varepsilon_{xx}\varepsilon_{yy}-\varepsilon_{xy}\varepsilon_{yx} &
=\varepsilon_{xx}\left(  \frac{\omega_{pi}^{2}}{\omega_{ci}^{2}}+\frac
{\omega_{pcr}^{2}}{\omega_{ccr}^{2}}\gamma_{cr0}\right)  -\varepsilon
_{xx}\frac{\omega_{pcr}^{2}}{\omega_{ccr}^{2}}\gamma_{cr0}^{2}c_{scr}^{2}%
\frac{\partial^{2}}{\partial x^{2}}\left(  \frac{\partial}{\partial t}\right)
^{-2}\\
&  -\varepsilon_{xx}\frac{\omega_{pi}^{2}}{\omega_{ci}^{2}}\frac{1}{m_{i}%
}\left(  T_{i0}+T_{e0}-\frac{G_{1}+G_{2}+G_{3}+G_{4}}{D}\frac{\partial
}{\partial t}\right)  \frac{\partial^{2}}{\partial x^{2}}\left(
\frac{\partial}{\partial t}\right)  ^{-2}\nonumber\\
&  +\frac{\omega_{pi}^{2}}{\omega_{ci}^{2}}\frac{\omega_{pcr}^{2}}%
{\omega_{ccr}^{2}}\gamma_{cr0}^{3}\left(  u_{cr}-u_{pl}\right)  ^{2}%
\frac{\partial^{2}}{\partial x^{2}}\left(  \frac{\partial}{\partial t}\right)
^{-2}.\nonumber
\end{align}
In these perturbations, we have $E_{1x}=0$ and $B_{1x,y}=0$, $B_{1z}\neq
0$.\textbf{\ }

\bigskip

\subsection{Dispersion relation}

After Fourier transformation of Equation (31) and substitution of Equation
(32), we derive the dispersion relation%
\begin{align}
\left(  1+\frac{\omega_{ci}^{2}}{\omega_{pi}^{2}}\frac{\omega_{pcr}^{2}%
}{\omega_{ccr}^{2}}\gamma_{cr0}\right)  \omega^{2}  &  =k_{x}^{2}c_{Ai}%
^{2}+\frac{\omega_{ci}^{2}}{\omega_{pi}^{2}}\frac{\omega_{pcr}^{2}}%
{\omega_{ccr}^{2}}\gamma_{cr0}^{2}k_{x}^{2}c_{scr}^{2}\\
&  +k_{x}^{2}\frac{1}{m_{i}}\left(  T_{i0}+T_{e0}+\frac{G_{1}+G_{2}%
+G_{3}+G_{4}}{D}i\omega\right) \nonumber\\
&  -\frac{1}{\varepsilon_{xx}}\frac{\omega_{pcr}^{2}}{\omega_{ccr}^{2}}%
\gamma_{cr0}^{3}k_{x}^{2}\left(  u_{cr}-u_{pl}\right)  ^{2}.\nonumber
\end{align}
Below, as above, we consider the streaming instability and influence of cosmic
rays on the thermal instability.

\bigskip

\subsubsection{\textit{Streaming instability}}

As above, we again neglect in the values $G_{i},i=1,2,3,4,$ and $D$ all the
frequencies $\Omega$. Then, Equation (33) takes the form%
\begin{equation}
\left(  1+\gamma_{cr0}^{-2}d\right)  \frac{\omega^{2}}{k_{x}^{2}}=-\frac
{d}{1+d}u_{cr}^{2}+c_{spl}^{2}+\gamma_{cr0}^{-1}dc_{scr}^{2}+c_{Ai}^{2},
\end{equation}
where we have omitted $u_{pl}$ in comparison with $u_{cr}$. This equation
describes an aperiodic instability, if the velocity of cosmic rays exceeds the
threshold given by Equation (26). The growth rate $\delta_{gr}$ when $u_{cr}$
exceeds $u_{crth}$ is the following:%
\begin{equation}
\delta_{gr}=\left[  \frac{d}{\left(  1+d\right)  \left(  1+\gamma_{cr0}%
^{-2}d\right)  }\right]  ^{1/2}k_{x}u_{cr}.
\end{equation}

\bigskip

\subsubsection{\textit{Thermal instability}}

Now, we take into account the contribution into Equation (33) of terms
describing the thermal instability in the fast thermal energy exchange regime
$\Omega_{\epsilon}\gg\partial/\partial t,\Omega_{Ti,e}$. The dispersion
relation becomes%
\begin{equation}
\frac{2\gamma\omega+i\Omega_{T,n}}{2\gamma\omega+i\gamma\Omega_{T}}%
=c_{spl}^{-2}\left[  \frac{d}{1+d}u_{cr}^{2}-\gamma_{cr0}^{-1}dc_{scr}%
^{2}-c_{Ai}^{2}+\left(  1+\gamma_{cr0}^{-2}d\right)  \frac{\omega^{2}}%
{k_{x}^{2}}\right]  .
\end{equation}
This equation is analogous to Equation (28). Depending on whether the
right-hand side of Equation (36) is much larger or smaller than unity, we will
have the Parker (1953) or the Field (1965) instability.

\bigskip

\section{Discussion and implications}

We first discuss cosmic-ray streaming instabilities found above, which are a
powerful source of magnetic amplification. The growth rates given by Equations
(27) and (35) have in somewhat a similar form and increase with decreasing of
the perturbation wavelength. The thresholds for the cases $k_{x}=0,$
$k_{y}\neq0$ and $k_{x}\neq0,$ $k_{y}=0$ are equal each other (see Equations
(24) at $u_{cr}\gg u_{pl}$ and (34)). Thus, streaming cosmic rays generate
perturbations in all directions across the ambient magnetic field. However, in
the case of strongly relativistic cosmic rays when $\gamma_{cr0}^{-2}d\gg1$
(the value $d$ is defined by Equation (25)), the growth rate given by Equation
(35) is $\gamma_{cr0}\gg1$ times larger than that described by Equation (27)
(for $k_{x}\sim k_{y}$). A spectrum of perturbations in the $\mathbf{k}$-space
is limited from above by Equations (15) and (16) and additional conditions
(see inequalities after Equations (18) and (30)). For the case $k_{x}%
=0,k_{y}\neq0$, Equation (15) of magnetization can be written in the "soft"
form%
\begin{equation}
\left(  \frac{\lambda_{y}}{2\pi}\right)  ^{2}\gtrsim\max\left\{  \frac{d}%
{1+d}\frac{u_{cr}^{2}}{\omega_{ci}^{2}};\frac{\gamma_{cr0}^{4}}{1+d}%
\frac{u_{cr}^{2}}{\omega_{ccr}^{2}}\right\}  ,
\end{equation}
where $\lambda_{y}$ ($\lambda_{x}$ below) is the wavelength along the $y$%
($x$)-direction. We have assumed that the threshold of instability is
exceeded. Equation (16) is the following: $1\gg k_{y}^{2}\rho_{i}^{2},$
$\gamma_{cr0}k_{y}^{2}\rho_{ccr}^{2}$. The "soft" Equation (15) for the case
$k_{x}\neq0,k_{y}=0$ is given by
\begin{equation}
\left(  \frac{\lambda_{x}}{2\pi}\right)  ^{2}\gtrsim\max\left\{  \frac
{d}{\left(  1+d\right)  \left(  1+\gamma_{cr0}^{-2}d\right)  }\frac{u_{cr}%
^{2}}{\omega_{ci}^{2}};\frac{\gamma_{cr0}^{4}d}{\left(  1+d\right)  \left(
1+\gamma_{cr0}^{-2}d\right)  }\frac{u_{cr}^{2}}{\omega_{ccr}^{2}}\right\}  .
\end{equation}
Equation (16) has the form $1\gg k_{x}^{2}\rho_{i}^{2},$ $\gamma_{cr0}%
^{3}k_{x}^{2}\rho_{ccr}^{2}$. Inequalities after Equations (18) and (30) are
satisfied. We see that the dependence of the right-hand sides of Equations
(37) and (38) on $d$ is different.

From Equations (28) and (36), it is followed that the relations between
magnetohydrodynamical parameters of thermal plasma and cosmic rays and the
perturbation wavelength determine the kind of thermal instability ranging from
the Parker (1953) to the Field (1965) type instability. Thus in our model, the
presence of streaming cosmic rays can only change the kind of thermal
instability, but not influence on its growth rates. This conclusion is
analogous to that in (Nekrasov \& Shadmehri 2012). However, the right-hand
sides of Equations (28) and (36) and in the corresponding equations of
(Nekrasov \& Shadmehri 2012) are quite different.

Let us now compare the growth rate for the streaming instability along the
background magnetic field found by Nekrasov \& Shadmehri (2012) with the
growth rates obtained in this paper. The growth rates given by Equations (27)
and (35) are of the same order of magnitude, if $\gamma_{cr0}\sim1$ or
$\gamma_{cr0}\gg1$ and $d\lesssim1$ (for the same wave numbers). In the case
$\gamma_{cr0}\gg1$ and $d\gg1$, the growth rate (35) is larger. Therefore, we
use Equation (35) for a comparison. The maximal growth rate in (Nekrasov \&
Shadmehri 2012) is equal to%
\[
\delta_{m}=2j_{cr0}\left(  \frac{\pi}{m_{cr}n_{cr0}c^{2}}\right)
^{1/2}\left(  \frac{\gamma_{cr0}^{-1}c_{A}^{2}}{\gamma_{cr0}^{-1}c_{scr}%
^{2}+c_{A}^{2}}\right)  ^{1/2},
\]
where $c_{A}=c_{Ai}\left(  1+\gamma_{cr0}^{-2}d\right)  ^{-1/2}$. The ratio of
this growth rate to the growth rate (35) for the same cosmic-ray drift
velocities is the following:%
\begin{equation}
\frac{\delta_{m}}{\delta_{gr}}=\left(  1+d^{-1}\right)  ^{1/2}\frac{c_{Ai}%
}{\left(  c_{scr}^{2}+\gamma_{cr0}c_{A}^{2}\right)  ^{1/2}}\frac{\omega_{pcr}%
}{k_{x}c}.
\end{equation}
Let cosmic rays be the protons. We estimate $k_{x}=k_{x\max}$ from Equation
(38)%
\[
k_{x\max}\approx\gamma_{cr0}^{-2}\left(  1+d^{-1}\right)  ^{1/2}\left(
1+\gamma_{cr0}^{-2}d\right)  ^{1/2}\frac{\omega_{ccr}}{u_{cr}}.
\]
Substituting this estimation into Equation (39) for the case $\gamma
_{cr0}c_{A}^{2}\gtrsim c_{scr}^{2}$, we obtain%
\begin{equation}
\frac{\delta_{m}}{\delta_{gr}}\approx\gamma_{cr0}^{3/2}\left(  \frac{n_{cr0}%
}{n_{i0}}\right)  ^{1/2}\frac{u_{cr}}{c_{Ai}}.
\end{equation}
Depending on parameters $u_{cr}$, $n_{cr0}$ and $B_{0}$, this relation can be
both less and larger then unity. In the opposite case, $\gamma_{cr0}c_{A}%
^{2}\ll c_{scr}^{2}$, Equation (39) takes the form
\begin{equation}
\frac{\delta_{m}}{\delta_{gr}}\approx\gamma_{cr0}^{3/2}\left(  \frac{n_{cr0}%
}{n_{i0}}\right)  ^{1/2}\frac{u_{cr}}{c_{Ai}}\frac{\gamma_{cr0}^{1/2}c_{A}%
}{c_{scr}}.
\end{equation}
We see that in this case the right-hand side of Equation (41) is smaller than
that of Equation (40). Thus, transverse streaming instabilities induced by the
cosmic-ray back-reaction can considerably contribute to turbulence of
astrophysical objects and amplification of magnetic fields.

We now consider some specific values of the growth rates (27) and (35) for
cosmic-ray protons in galaxy clusters. For the ICM, we take $T_{0}=3$ keV and
$B_{0}=1$ $\mu$G. Then, we obtain $\omega_{ci}\simeq10^{-2}$ s$^{-1}$ and
$c_{spl}=10^{8}$ cm s$^{-1}$. In the case of weakly relativistic cosmic rays,
$u_{cr}\sim c$ and $\gamma_{cr0}\sim1$, the parameter $d\sim n_{cr0}/n_{i0}%
\ll1$. Since $u_{cr}\gg c_{spl}$, the wave number $k_{y}$ is less than
$\omega_{ci}/u_{cr}\sim$ $3.3\times10^{-13}$ cm$^{-1}$ (see Equation (37)). In
the real case in which $n_{cr0}/n_{i0}\ll c_{spl}^{2}/u_{cr}^{2}$, or
$n_{cr0}/n_{i0}\ll10^{-5}$, the wave number $k_{x}$ is limited from above by
$\omega_{ci}/c_{spl}$, or $10^{-10}$ cm$^{-1}$. Thus, the upper estimations of
the growth rates (27) and (35) are $\delta_{gr}\sim\omega_{ci}\left(
n_{cr0}/n_{i0}\right)  ^{1/2}$ and $\delta_{gr}=\omega_{ci}\left(
n_{cr0}/n_{i0}\right)  ^{1/2}\left(  u_{cr}/c_{spl}\right)  $, respectively.
These values are considerably larger than the Bell instability (Bell 2004). In
the case of ultrarelativistic cosmic rays when $d\gg1$ but $\gamma_{cr0}%
^{-2}d\ll1$, or $\gamma_{cr0}^{-3}$ $\ll n_{cr0}/n_{i0}\ll\gamma_{cr0}^{-1}$,
we obtain $k_{y}\lesssim\left(  n_{cr0}/\gamma_{cr0}n_{i0}\right)
^{1/2}\left(  \omega_{ci}/u_{cr}\right)  $ and $k_{x}\lesssim\omega
_{ci}/\gamma_{cr0}^{2}u_{cr} $ (see Equations (37) and (38)). Correspondingly,
the limiting growth rates (27) and (35) are the same and equal to $\delta
_{gr}\sim\gamma_{cr0}^{-2}\omega_{ci}$. We note that the last expression is
independent from the density of cosmic rays. In the case $\gamma_{cr0}%
^{-2}d\gg1$, or $n_{cr0}/n_{i0}\gg\gamma_{cr0}^{-1}$, the region of
wavelengths of unstable perturbations in the $y$-direction and the growth rate
remain the same as for the case $\gamma_{cr0}^{-2}d\ll1$. The wave numbers of
the $x$-perturbations satisfy to $k_{x}\lesssim\left(  n_{cr0}/\gamma
_{cr0}^{3}n_{i0}\right)  ^{1/2}\left(  \omega_{ci}/u_{cr}\right)  $ and the
corresponding growth rate is equal to $\delta_{gr}\sim\gamma_{cr0}^{-2}%
\omega_{ci}$ as above. In the case $d\sim1$, or $\ n_{cr0}/n_{i0}\sim
\gamma_{cr0}^{-3}$, the growth rates are the same in both cases and are equal
to the last expression.

In Section 3, we have marked some other mechanisms of the appearance of
perpendicular currents except of the model by Riquelme \& Spitkovsky (2010).
In general, the results for new instabilities obtained here do not change
because we use in the equations of motion for species the cosmic-ray and
plasma drift velocities\textbf{ }$\mathbf{u}_{cr}$ and\textbf{ }%
$\mathbf{u}_{pl}$, which are not specified for concrete mechanisms. The
contribution of electron drifts will be negligible. The specific\textbf{
}forms\textbf{ }of values\textbf{ }$\mathbf{u}_{cr}$ and\textbf{ }%
$\mathbf{u}_{pl}$ will depend on the origin of the perpendicular current. In
the case $\mathbf{u}_{cr}\gg\mathbf{u}_{pl}$, the return current of a plasma
in dispersion relations (22) and (33) is not important. \ 

In this paper, we for simplicity did not take into account the electron-ion
collisions in the momentum equation (1). For perturbations across the magnetic
field, the condition allowing to neglect this effect can be found in (Nekrasov
2012, Equation (34)) and has the form%
\[
1\gg\frac{\nu_{ie}\omega}{\omega_{ci}^{2}}\left(  1+\frac{k_{x,y}^{2}%
c_{spl}^{2}}{\omega^{2}}\right)  .
\]
The presence of the ion drift velocity $u_{pl}$ resulting in the Doppler shift
(see Appendix A) does not influence on this condition because $\omega\gg
k_{y}u_{pl}$ for the streaming instability and $\omega\gtrsim k_{y}u_{pl}$ for
the thermal one (see Section 5.2). However, the collision frequency in the
energy equation, $\Omega_{\epsilon}=2\nu_{ie}$, is added to
frequencies\ $\partial/\partial t$ and $\Omega_{Ti,e}$ (see Equations (A29)
and (A30)). Thus, the contributions to the dispersion relation of collisions
in the momentum equation and in the energy equation are quite different.
Therefore, the collisional energy exchange between electrons and ions is
included in our analysis.

The model explored here with cosmic rays propagating across the ambient
magnetic field has been considered by Bell (2005) (a general case) and have
been applied by Riquelme \& Spitkovsky (2010) for the problem of the magnetic
field amplification in the upstream region of the supernova remnant shocks.
However, streaming cosmic-ray driven instabilities can exist in a variety of
environments. Therefore, we believe, wherever there is a cosmic-ray streaming,
these instabilities may play a significant role. For example, the model
described above can be applied to the ICM where cosmic rays are an important
ingredient (Loewenstein et al. 1991; Guo \& Oh 2008; Sharma et al. 2009;
Sharma et al. 2010). Observations show that many cavities or bubbles in the
ICM contain cosmic rays and magnetic fields (e.g., Guo \& Oh 2008). A
substantial amount of cosmic rays may escape from these buoyantly rising
bubbles (En\ss lin 2003), which could be disrupted by the Rayleigh-Taylor and
Kelvin-Helmholtz instabilities as they rise through the ICM (Fabian et al.
2006). Cosmic rays may also be produced by other processes near a central
active galactic nucleus of the galaxy cluster. Structure formation shocks,
merger shocks and supernovae may also inject cosmic rays into the ICM (e.g.,
Voelk et al. 1996; Berezinsky et al. 1997). Thus, various cosmic-ray streaming
instabilities considered in particular in this paper can be a powerful source
of generation of magnetic fields in astrophysical settings.

\bigskip

\section{Conclusion}

We have investigated streaming and thermal instabilities of astrophysical
plasmas consisting of electrons, ions, and cosmic rays propagating across the
background magnetic field. The drift velocity of cosmic rays can be
relativistic, however their mean energy is assumed to be small
(non-relativistic). The return current of the background plasma and the
back-reaction of magnetized cosmic rays are taken into account. We have
considered perturbations which are transverse to the background magnetic field
and are along and across the cosmic-ray drift velocity. The case of
perturbations along the magnetic field was treated by Nekrasov \& Shadmehri
(2012) where the growth rate due to the back-reaction of cosmic rays
considerably larger than that obtained by Bell (2004, 2005) and Riquelme \&
Spitkovsky (2010) has been found. In the present case, we have shown that for
sufficiently short-wavelength perturbations the growth rates obtained can in
their turn exceed the growth rate found in (Nekrasov \& Shadmehri 2012). This
new result increases the role of cosmic-ray streaming instabilities in
amplification of magnetic fields in astrophysical environments.

We have found that the thermal instability is not subject to the action of
cosmic rays in the model under consideration. The dispersion relations derived
for thermal instability include sound velocities of plasma and cosmic rays,
Alfv\'{e}n and cosmic-ray drift velocities. The relations between these
parameters determine the kind of thermal instability ranging from the Parker
(1953) to the Field (1965) type instability. However, the growth rates of
thermal instabilities do not change.

The results of this paper can be applied to investigations of weakly
collisional electron-ion astrophysical objects such as supernova remnant
shocks, galaxy clusters and others, which include the dynamics of streaming
cosmic rays.

\bigskip

\section*{References}

Achterberg, A. 1983, A\&A, 119, 274

Balbus, S. A., \& Soker, N. 1989, ApJ, 341, 611

Begelman, M. C., \& McKee, C. F. 1990, ApJ, 358, 375

Begelman, M. C., \& Zweibel, E. G. 1994, ApJ, 431, 689

Bell, A. R. 2004, MNRAS, 353, 550

Bell, A. R. 2005, MNRAS, 358, 181

Berezinsky, V. S., Blasi, P., \& Ptuskin, V. S. 1997, ApJ, 487, 529

Beresnyak, A., Jones, T. W., \& Lazarian, A. 2009, ApJ, 707, 1541

Berk, H. L., \& Pearlstein, L. D. 1976, Phys. Fluids, 19, 1831

Bogdanovi\'{c}, T., Reynolds, C. S., Balbus, S. A., \& Parrish, I. J. 2009,
ApJ, 704, 211

Braginskii, S. I. 1965, Rev. Plasma Phys., 1, 205

Cavagnolo, K. W., Donahue, M., Voil, G. M., \& Sun, M. 2008, ApJ, 683, L107

Conselice, C. J., Gallagher, J. S., III, \& Wyse, R. F. G. 2001, AJ, 122, 2281

Cox, J. L. Jr, \& Bennett, W. H. 1970, Phys. Fluids, 13, 182

Dzhavakhishvili, D. I., \& Tsintsadze, N. L. 1973, Sov. Phys. JEPT, 37, 666

En\ss lin, T. A. 2003, A\&A, 399, 409

En\ss lin, T., Pfrommer, C., Miniati, F., \& Subramanian, K. 2011, A\&A, 527, A99

Everett, J. E., Zweibel, E. G., Benjamin, R. A., McCammon, D., Rocks, L., \&
Gallagher, J. S., III. 2008, ApJ, 674, 258

Fabian, A. C., Sanders, J. S., Taylor, G. B., Allen, S. W., Crawford, C. S.,
Johnstone, R. M., \& Iwasawa, K. 2006, MNRAS, 366, 417

Ferland, G. J., Fabian, A. C., Hatch, N. A., Johnstone, R. M., Porter, R. L.,
van Hoof, P. A. M., \& Williams, R. J. R. 2009, MNRAS, 392, 1475

Field, G. B. 1965, ApJ, 142, 531

Field, G. B., Goldsmith, D. W., \& Habing, H. J. 1969, ApJ, 155, L149

Fukue, T., \& Kamaya, H. 2007, ApJ, 669, 363

Gammie, C. F. 1996, ApJ, 457, 355

Gratton, F. T., Gnavi, G., Galv\~{a}o, R. M. O., \& Gomberoff, L. 1998,
Ap\&SS, 256, 311

Guo, F., \& Oh, S. P. 2008, MNRAS, 384, 251

Haim, L. 2009, Nonlinear waves and shocks in relativistic two-fluid
hydrodynamics, \textit{Thesis for the degree of master of science}, Department
of Physics, Faculty of Natural Sciences, Ben-Gurion University of the Negev

Hammer, D. A., \& Rostoker, N. 1970, Phys. Fluids, 13, 1831

Hazeltine, R. D., \& Mahajan, S. M. 2000, Relativistic magnetohydrodynamics,
Institute for Fusion Studies (Austin: Univ. of Texas); http://w3fusion.ph.utexas.edu/ifs/ifsreports/913\_hazeltine.pdf

Hennebelle, P., \& P\'{e}rault, M. 2000, A\&A, 359, 1124

Inoue, T., \& Inutsuka, S. 2008, ApJ, 687, 303

Koyama, H., \& Inutsuka, S. 2000, ApJ, 532, 980

Kulsrud, R., \& Pearce, W. P. 1969, ApJ, 156, 445

Loewenstein, M. 1990, ApJ, 349, 471

Loewenstein, M., Zweibel, E. G., \& Begelman, M. C. 1991, ApJ, 377, 392

Lontano, M., Bulanov, S., \& Koga, J. 2001, Phys. Plasmas, 8, 5113

Mathews, W., \& Bregman, J. 1978, ApJ, 224, 308

Mikhailovskii, A. B., Onischenko, O. G., \& Tatarinov, E. G. 1985, Plasma
Phys. Control. Fusion, 27, 527

Mofiz, U. A., \& Khan, F. 1993, Ap\&SS, 201, 53

Nekrasov, A. K. 2007, Phys. Plasmas, 14, 062107

Nekrasov, A. K. 2011, ApJ, 739, A88

Nekrasov, A. K. 2012, MNRAS, 419, 522

Nekrasov, A. K. 2013, Plasma Phys. Control. Fusion, 55, 085007

Nekrasov, A. K., \& Shadmehri, M. 2012, ApJ, 756, A77

O'Dea, C. P., Baum, S. A., Privon, G., et al. 2008, ApJ, 681, 1035

Parker, E. N. 1953, ApJ, 117, 431

Parrish, I. J., Quataert, E., \& Sharma, P. 2009, ApJ, 703, 96

Riquelme, M. A., \& Spitkovsky, A. 2009, ApJ, 694, 626

Riquelme, M. A., \& Spitkovsky, A. 2010, ApJ, 717, 1054

Roberts, T. G., \& Bennett, W. H. 1968, Plasma Phys., 10, 381

Sakai, J., \& Kawata, T. J. 1980, Phys. Soc. Japan, 49, 747\ 

Salom\'{e}, P., Combes, F., Edge, A. C., et al. 2006, A\&A, 454, 437

Samui, S., Subramanian, K., \& Srianand, R. 2010, MNRAS, 402, 2778

S\'{a}nchez-Salcedo, F. J., V\'{a}zquez-Semadeni, E., \& Gazol, A. 2002, ApJ,
577, 768

Shadmehri, M., Nejad-Asghar, M., \& Khesali, A. 2010, Ap\&SS, 326, 83

Sharma, P., Chandran, B. D. G., Quataert, E., \& Parrish, I. J. 2009, ApJ,
699, 348

Sharma, P., Parrish, I. J., \& Quataert, E. 2010, ApJ, 720, 652

Skilling, J. 1975, MNRAS, 172, 557

Toepfer, A. J. 1971, Phys. Rev. A, 3, 1444

Tozzi, P., \& Norman, C. 2001, ApJ, 546, 63

V\'{a}zquez-Semadeni, E., Ryu, D., Passot, T., Gonz\'{a}lez, R. F., \& Gazol,
A. 2006, ApJ, 643, 245

Voelk, H. J., Aharonian, F. A., \& Breitschwerdt, D. 1996, Space Sci. Rev.,
75, 279

Wallis, G., Sauer, K., S\"{u}nder, D., Rosinskii, S. E., Rukhadze, A. A., \&
Rukhlin, V. G. 1975, Sov. Phys. Usp., 17, 492

Yusef-Zadeh, F., Wardle, M., \& Roy, S. 2007, ApJ, 665, L123

Zweibel, E. G. 2003, ApJ, 587, 625

Zweibel E. G., \& Everett, J. E. 2010, ApJ, 709, 1412

\bigskip

\begin{appendix}
\section{Appendix}
\bigskip
\subsection{Perturbed velocities of ions and electrons}
We put in Equation (1) $\mathbf{v}_{j}=\mathbf{v}_{j0}+\mathbf{v}_{j1}$,
$p_{j}=p_{j0}+p_{j1}$, $\mathbf{E=E}_{0}+\mathbf{E}_{1}$, $\mathbf{B=B}%
_{0}+\mathbf{B}_{1}$, where the subscript $0$ denotes equilibrium uniform
parameters and the subscript $1$ relates to perturbations. Then the linearized
version of this equation takes the form%
\begin{equation}
\frac{\partial\mathbf{v}_{j1}}{\partial t}+\mathbf{v}_{j0}\cdot\mathbf{\nabla
v}_{j1}=-\frac{\mathbf{\nabla}T_{j1}}{m_{j}}-\frac{T_{j0}}{m_{j}}%
\frac{\mathbf{\nabla}n_{j1}}{n_{j0}}+\mathbf{F}_{j1}\mathbf{+}\frac{q_{j}%
}{m_{j}c}\mathbf{v}_{j1}\times\mathbf{B}_{0},\tag{A1}%
\end{equation}
where we have used that $p_{j1}=n_{j0}T_{j1}+n_{j1}T_{j0}$ ($n_{j}%
=n_{j0}+n_{j1}$, $T_{j}=T_{j0}+T_{j1}$) and introduced notation%
\begin{equation}
\mathbf{F}_{j1}=\frac{q_{j}}{m_{j}}\mathbf{E}_{1}\mathbf{+}\frac{q_{j}}%
{m_{j}c}\mathbf{v}_{j0}\times\mathbf{B}_{1}.\tag{A2}%
\end{equation}
From Equation (A1), we find expressions for the ion velocities $v_{i1x,y}$ in
the form%
\begin{align}
\Omega_{i}^{2}v_{i1x} &  =\frac{1}{m_{i}}L_{ix}T_{i1}-\frac{T_{i0}}{m_{i}%
}L_{ix}\left(  \frac{\partial}{\partial t}+v_{i0y}\frac{\partial}{\partial
y}\right)  ^{-1}\mathbf{\nabla}\cdot\mathbf{v}_{i1}\tag{A3}\\
&  \mathbf{+}\omega_{ci}F_{i1y}+\left(  \frac{\partial}{\partial t}%
+v_{i0y}\frac{\partial}{\partial y}\right)  F_{i1x}\nonumber
\end{align}
and
\begin{align}
\Omega_{i}^{2}v_{i1y} &  =\frac{1}{m_{i}}L_{iy}T_{i1}-\frac{T_{i0}}{m_{i}%
}L_{iy}\left(  \frac{\partial}{\partial t}+v_{i0y}\frac{\partial}{\partial
y}\right)  ^{-1}\mathbf{\nabla}\cdot\mathbf{v}_{i1}\tag{A4}\\
&  -\omega_{ci}F_{i1x}+\left(  \frac{\partial}{\partial t}+v_{i0y}%
\frac{\partial}{\partial y}\right)  F_{i1y}.\nonumber
\end{align}
In Equations (A3) and (A4), we have used the linearized continuity equation
(2). The following notations are here introduced:%
\begin{align}
\Omega_{i}^{2} &  =\left(  \frac{\partial}{\partial t}+v_{i0y}\frac{\partial
}{\partial y}\right)  ^{2}+\omega_{ci}^{2},\tag{A5}\\
L_{ix} &  =-\omega_{ci}\frac{\partial}{\partial y}-\left(  \frac{\partial
}{\partial t}+v_{i0y}\frac{\partial}{\partial y}\right)  \frac{\partial
}{\partial x},\nonumber\\
L_{iy} &  =\omega_{ci}\frac{\partial}{\partial x}-\left(  \frac{\partial
}{\partial t}+v_{i0y}\frac{\partial}{\partial y}\right)  \frac{\partial
}{\partial y}.\nonumber
\end{align}
Analogous equations for the electrons are the following:%
\begin{equation}
\Omega_{e}^{2}v_{e1x}=\frac{1}{m_{e}}L_{ex}T_{e1}-\frac{T_{e0}}{m_{e}}%
L_{ex}\left(  \frac{\partial}{\partial t}\right)  ^{-1}\mathbf{\nabla}%
\cdot\mathbf{v}_{e1}+\omega_{ce}F_{e1y}+\frac{\partial F_{e1x}}{\partial
t},\tag{A6}%
\end{equation}%
\begin{equation}
\Omega_{e}^{2}v_{e1y}=\frac{1}{m_{e}}L_{ey}T_{e1}-\frac{T_{e0}}{m_{e}}%
L_{ey}\left(  \frac{\partial}{\partial t}\right)  ^{-1}\mathbf{\nabla}%
\cdot\mathbf{v}_{e1}-\omega_{ce}F_{e1x}+\frac{\partial F_{e1y}}{\partial
t},\tag{A7}%
\end{equation}
where
\begin{align}
\Omega_{e}^{2} &  =\frac{\partial^{2}}{\partial t^{2}}+\omega_{ce}%
^{2},\tag{A8}\\
L_{ex} &  =-\omega_{ce}\frac{\partial}{\partial y}-\frac{\partial^{2}%
}{\partial x\partial t},\nonumber\\
L_{ey} &  =\omega_{ce}\frac{\partial}{\partial x}-\frac{\partial^{2}}{\partial
y\partial t}.\nonumber
\end{align}
We do not consider the longitudinal velocity $v_{j1z}$ because as can be shown
in the case $\partial/\partial z=0$ this velocity only depends on the electric
field $E_{1z}$, $\partial v_{j1z}/\partial t=\left(  q_{j}/m_{j}\right)
E_{1z}$, and the transverse and longitudinal wave equations are split.
\bigskip
\subsection{Perturbed temperatures of ions and electrons}
We find now equations for the temperature perturbations $T_{i,e1}$. We here
assume that equilibrium temperatures $T_{i0}$ and $T_{e0}$ are equal one
another, $T_{i0}=T_{e0}=T_{0}$. The case $T_{i0}\neq T_{e0}$ for thermal
instability has been considered by Nekrasov (2011, 2012). For equal
temperatures, the terms connected with the perturbation of thermal energy
exchange frequency in Equations (3) and (4) will be absent. However for
convenience of calculations, we formally retain different notations for the
ion and electron temperatures. From Equations (3) and (4) in the linear form,
we obtain equations for the temperature perturbations
\begin{equation}
D_{1i}T_{i1}-D_{2i}T_{e1}=C_{1i}\mathbf{\nabla}\cdot\mathbf{v}_{i1},\tag{A9}%
\end{equation}%
\begin{equation}
D_{1e}T_{e1}-D_{2e}T_{i1}=C_{1e}\mathbf{\nabla}\cdot\mathbf{v}_{e1},\tag{A10}%
\end{equation}
where notations are introduced%
\begin{align}
D_{1i} &  =\left[  \left(  \frac{\partial}{\partial t}+v_{i0y}\frac{\partial
}{\partial y}\right)  +\Omega_{Ti}+\Omega_{ie}\right]  \left(  \frac{\partial
}{\partial t}+v_{i0y}\frac{\partial}{\partial y}\right)  ,\tag{A11}\\
D_{2i} &  =\Omega_{ie}\left(  \frac{\partial}{\partial t}+v_{i0y}%
\frac{\partial}{\partial y}\right)  ,\nonumber\\
C_{1i} &  =T_{i0}\left[  -\left(  \gamma-1\right)  \left(  \frac{\partial
}{\partial t}+v_{i0y}\frac{\partial}{\partial y}\right)  +\Omega_{ni}\right]
,\nonumber\\
D_{1e} &  =\left(  \frac{\partial}{\partial t}+\Omega_{Te}+\Omega_{ei}\right)
\frac{\partial}{\partial t},\nonumber\\
D_{2e} &  =\Omega_{ei}\frac{\partial}{\partial t},\nonumber\\
C_{1e} &  =T_{e0}\left[  -\left(  \gamma-1\right)  \frac{\partial}{\partial
t}+\Omega_{ne}\right]  .\nonumber
\end{align}
For obtaining Equations (A9) and (A10), we have used Equations (2) and (12).
The frequencies in Equation (A11) are the following:%
\begin{align}
\Omega_{Tj} &  =\left(  \gamma-1\right)  \frac{\partial%
\mathcal{L}%
_{j}\left(  n_{j0},T_{j0}\right)  }{n_{j0}\partial T_{j0}},\Omega_{nj}=\left(
\gamma-1\right)  \frac{\partial%
\mathcal{L}%
_{j}\left(  n_{j0},T_{j0}\right)  }{T_{j0}\partial n_{j0}},\tag{A12}\\
\Omega_{ie} &  =\nu_{ie}^{\varepsilon}\left(  n_{e0},T_{e0}\right)
,\Omega_{ei}=\nu_{ei}^{\varepsilon}\left(  n_{i0},T_{e0}\right)  .\nonumber
\end{align}
From Equations (A9) and (A10), we find equations for $T_{i1}$ and $T_{e1}$%
\begin{equation}
DT_{i1}=G_{4}\mathbf{\nabla}\cdot\mathbf{v}_{i1}+G_{3}\mathbf{\nabla}%
\cdot\mathbf{v}_{e1}\tag{A13}%
\end{equation}
and%
\begin{equation}
DT_{e1}=G_{1}\mathbf{\nabla}\cdot\mathbf{v}_{e1}+G_{2}\mathbf{\nabla}%
\cdot\mathbf{v}_{i1}.\tag{A14}%
\end{equation}
Here, we have%
\begin{align}
D &  =D_{1i}D_{1e}-D_{2i}D_{2e},\tag{A15}\\
G_{1} &  =D_{1i}C_{1e},G_{2}=D_{2e}C_{1i},\nonumber\\
G_{3} &  =D_{2i}C_{1e},G_{4}=D_{1e}C_{1i}.\nonumber
\end{align}
\bigskip
\subsection{Expressions for $\mathbf{\nabla\cdot v}_{i,e1}$}
We now substitute temperature perturbations $T_{i,e1}$ defined by Equations
(A13) and (A14) into Equations (A3) and (A4). Then applying operators
$\partial/\partial x$ and $\partial/\partial y$ to Equations (A3) and (A4),
respectively, and adding them, we find equation for $\mathbf{\nabla}%
\cdot\mathbf{v}_{i1}$%
\begin{equation}
L_{1i}\mathbf{\nabla}\cdot\mathbf{v}_{i1}=-L_{2i}\mathbf{\nabla}%
\cdot\mathbf{v}_{e1}+\Phi_{i1},\tag{A16}%
\end{equation}
where
\begin{align}
L_{1i} &  =\Omega_{i}^{2}+\frac{1}{m_{i}}\left[  \frac{G_{4}}{D}\left(
\frac{\partial}{\partial t}+v_{i0y}\frac{\partial}{\partial y}\right)
-T_{i0}\right]  \mathbf{\nabla}^{2},\tag{A17}\\
L_{2i} &  =\frac{1}{m_{i}}\frac{G_{3}}{D}\left(  \frac{\partial}{\partial
t}+v_{i0y}\frac{\partial}{\partial y}\right)  \mathbf{\nabla}^{2},\nonumber\\
\Phi_{i1} &  =\omega_{ci}\left(  \frac{\partial F_{i1y}}{\partial x}%
-\frac{\partial F_{i1x}}{\partial y}\right)  +\left(  \frac{\partial}{\partial
t}+v_{i0y}\frac{\partial}{\partial y}\right)  \mathbf{\nabla\cdot F}%
_{i1}.\nonumber
\end{align}
Analogously, using Equations (A6) and (A7), we obtain
\begin{equation}
L_{1e}\mathbf{\nabla}\cdot\mathbf{v}_{e1}=-L_{2e}\mathbf{\nabla}%
\cdot\mathbf{v}_{i1}+\Phi_{e1},\tag{A18}%
\end{equation}
where%
\begin{align}
L_{1e} &  =\Omega_{e}^{2}+\frac{1}{m_{e}}\left(  \frac{G_{1}}{D}\frac
{\partial}{\partial t}-T_{e0}\right)  \mathbf{\nabla}^{2},\tag{A19}\\
L_{2e} &  =\frac{1}{m_{e}}\frac{G_{2}}{D}\frac{\partial}{\partial
t}\mathbf{\nabla}^{2},\nonumber\\
\Phi_{e1} &  =\omega_{ce}\left(  \frac{\partial F_{e1y}}{\partial x}%
-\frac{\partial F_{e1x}}{\partial y}\right)  +\frac{\partial}{\partial
t}\mathbf{\nabla\cdot F}_{e1}.\nonumber
\end{align}
From Equations (A16) and (A18), we find
\begin{equation}
L\mathbf{\nabla}\cdot\mathbf{v}_{i1}=L_{1e}\Phi_{i1}-L_{2i}\Phi_{e1}\tag{A20}%
\end{equation}
and%
\begin{equation}
L\mathbf{\nabla}\cdot\mathbf{v}_{e1}=L_{1i}\Phi_{e1}-L_{2e}\Phi_{i1}.\tag{A21}%
\end{equation}
The operator $L$ is given by%
\begin{equation}
L=L_{1i}L_{1e}-L_{2i}L_{2e}.\tag{A22}%
\end{equation}
\bigskip
\subsection{Equations for ion and electron velocities via\textit{\ }%
$\mathbf{F}_{i,e1}$}
Using Equations (A3), (A4), (A13), (A20), and (A21), we obtain the following
equations for components of the perturbed ion velocity:%
\begin{equation}
\Omega_{i}^{2}v_{i1x}=\frac{L_{ix}}{m_{i}DL}\left(  A_{1i}\Phi_{i1}-A_{2i}%
\Phi_{e1}\right)  +\omega_{ci}F_{i1y}+\left(  \frac{\partial}{\partial
t}+v_{i0y}\frac{\partial}{\partial y}\right)  F_{i1x}\tag{A23}%
\end{equation}
and%
\begin{equation}
\Omega_{i}^{2}v_{i1y}=\frac{L_{iy}}{m_{i}DL}\left(  A_{1i}\Phi_{i1}-A_{2i}%
\Phi_{e1}\right)  -\omega_{ci}F_{i1x}+\left(  \frac{\partial}{\partial
t}+v_{i0y}\frac{\partial}{\partial y}\right)  F_{i1y}.\tag{A24}%
\end{equation}
The operators $A_{1,2i}$ are given by%
\begin{align}
A_{1i} &  =\left[  G_{4}-DT_{i0}\left(  \frac{\partial}{\partial t}%
+v_{i0y}\frac{\partial}{\partial y}\right)  ^{-1}\right]  L_{1e}-G_{3}%
L_{2e},\tag{A25}\\
A_{2i} &  =\left[  G_{4}-DT_{i0}\left(  \frac{\partial}{\partial t}%
+v_{i0y}\frac{\partial}{\partial y}\right)  ^{-1}\right]  L_{2i}-G_{3}%
L_{1i}.\nonumber
\end{align}
Equations for components of the perturbed electron velocity are found by using
Equations (A6), (A7), (A14), (A20), and (A21)%
\begin{equation}
\Omega_{e}^{2}v_{e1x}=\frac{L_{ex}}{m_{e}DL}\left(  A_{1e}\Phi_{e1}-A_{2e}%
\Phi_{i1}\right)  +\omega_{ce}F_{e1y}+\frac{\partial F_{e1x}}{\partial
t},\tag{A26}%
\end{equation}%
\begin{equation}
\Omega_{e}^{2}v_{e1y}=\frac{L_{ey}}{m_{e}DL}\left(  A_{1e}\Phi_{e1}-A_{2e}%
\Phi_{i1}\right)  -\omega_{ce}F_{e1x}+\frac{\partial F_{e1y}}{\partial
t}.\tag{A27}%
\end{equation}
Here,%
\begin{align}
A_{1e} &  =\left[  G_{1}-DT_{e0}\left(  \frac{\partial}{\partial t}\right)
^{-1}\right]  L_{1i}-G_{2}L_{2i},\tag{A28}\\
A_{2e} &  =\left[  G_{1}-DT_{e0}\left(  \frac{\partial}{\partial t}\right)
^{-1}\right]  L_{2e}-G_{2}L_{1e}.\nonumber
\end{align}
\bigskip
\subsection{Expressions for $D$ and $G_{1,2,3,4}$}
We now give expressions for $D$ and $G_{1,2,3,4}$ defined by Equation (A15).
Using Equation (A11), we find%
\begin{align}
\left(  \frac{\partial}{\partial t}+v_{i0y}\frac{\partial}{\partial y}\right)
^{-1}\left(  \frac{\partial}{\partial t}\right)  ^{-1}D &  =\left[  \left(
\frac{\partial}{\partial t}+v_{i0y}\frac{\partial}{\partial y}\right)
+\Omega_{Ti}\right]  \left(  \frac{\partial}{\partial t}+\Omega_{Te}\right)
\tag{A29}\\
&  +\left(  \frac{\partial}{\partial t}+\Omega_{Te}\right)  \Omega
_{ie}+\left[  \left(  \frac{\partial}{\partial t}+v_{i0y}\frac{\partial
}{\partial y}\right)  +\Omega_{Ti}\right]  \Omega_{ei}\nonumber
\end{align}
and%
\begin{align}
G_{1} &  =T_{e0}\left[  \left(  \frac{\partial}{\partial t}+v_{i0y}%
\frac{\partial}{\partial y}\right)  +\Omega_{Ti}+\Omega_{ie}\right]  \left[
-\left(  \gamma-1\right)  \frac{\partial}{\partial t}+\Omega_{ne}\right]
\left(  \frac{\partial}{\partial t}+v_{i0y}\frac{\partial}{\partial y}\right)
,\tag{A30}\\
G_{2} &  =T_{i0}\Omega_{ei}\left[  -\left(  \gamma-1\right)  \left(
\frac{\partial}{\partial t}+v_{i0y}\frac{\partial}{\partial y}\right)
+\Omega_{ni}\right]  \frac{\partial}{\partial t},\nonumber\\
G_{3} &  =T_{e0}\Omega_{ie}\left[  -\left(  \gamma-1\right)  \frac{\partial
}{\partial t}+\Omega_{ne}\right]  \left(  \frac{\partial}{\partial t}%
+v_{i0y}\frac{\partial}{\partial y}\right)  ,\nonumber\\
G_{4} &  =T_{i0}\left(  \frac{\partial}{\partial t}+\Omega_{Te}+\Omega
_{ei}\right)  \left[  -\left(  \gamma-1\right)  \left(  \frac{\partial
}{\partial t}+v_{i0y}\frac{\partial}{\partial y}\right)  +\Omega_{ni}\right]
\frac{\partial}{\partial t}.\nonumber
\end{align}
\bigskip
\subsection{Simplified expressions for $A_{1,2i}$ and $A_{1,2e}$}
We can further simplify expressions for $A_{1,2i}$ and $A_{1,2e}$ given by
Equations (A25) and (A28). Using Equation (A17), we obtain%
\begin{equation}
A_{2i}=-G_{3}\Omega_{i}^{2}.\tag{A31}%
\end{equation}
The expression for $A_{1i}$ can be given in the form%
\begin{equation}
A_{1i}=\left[  G_{4}-DT_{i0}\left(  \frac{\partial}{\partial t}+v_{i0y}%
\frac{\partial}{\partial y}\right)  ^{-1}\right]  \Omega_{e}^{2}-\frac
{1}{m_{e}}\mathbf{\nabla}^{2}K,\tag{A32}%
\end{equation}
where we have used Equation (A19). The following notation is introduced in
Equation (A32):%
\begin{equation}
K=\frac{1}{D}\left(  G_{2}G_{3}-G_{1}G_{4}\right)  \frac{\partial}{\partial
t}+G_{4}T_{e0}+G_{1}T_{i0}\left(  \frac{\partial}{\partial t}+v_{i0y}%
\frac{\partial}{\partial y}\right)  ^{-1}\frac{\partial}{\partial t}%
-DT_{i0}T_{e0}\left(  \frac{\partial}{\partial t}+v_{i0y}\frac{\partial
}{\partial y}\right)  ^{-1}.\tag{A33}%
\end{equation}
Analogously, we will have%
\begin{equation}
A_{2e}=-G_{2}\Omega_{e}^{2}\tag{A34}%
\end{equation}
and%
\begin{equation}
A_{1e}=\left[  G_{1}-DT_{e0}\left(  \frac{\partial}{\partial t}\right)
^{-1}\right]  \Omega_{i}^{2}-\frac{1}{m_{i}}\mathbf{\nabla}^{2}\left(
\frac{\partial}{\partial t}+v_{i0y}\frac{\partial}{\partial y}\right)  \left(
\frac{\partial}{\partial t}\right)  ^{-1}K.\tag{A35}%
\end{equation}
Calculations show that the value $D^{-1}\left(  G_{2}G_{3}-G_{1}G_{4}\right)
$ takes the simple form%
\begin{equation}
\frac{1}{D}\left(  G_{2}G_{3}-G_{1}G_{4}\right)  =-T_{i0}T_{e0}\left[
-\left(  \gamma-1\right)  \left(  \frac{\partial}{\partial t}+v_{i0y}%
\frac{\partial}{\partial y}\right)  +\Omega_{ni}\right]  \left[  -\left(
\gamma-1\right)  \frac{\partial}{\partial t}+\Omega_{ne}\right]  .\tag{A36}%
\end{equation}
Using Equations (A29), (A30), and (A36), we can also rewrite the value $K$
defined by Equation (A33) in the simple form%
\begin{equation}
K=-T_{i0}T_{e0}\left(  W_{i}W_{e}+W_{i}\Omega_{ei}+W_{e}\Omega_{ie}\right)
\frac{\partial}{\partial t}.\tag{A37}%
\end{equation}
Here, notations are introduced
\begin{align}
W_{i} &  =\gamma\left(  \frac{\partial}{\partial t}+v_{i0y}\frac{\partial
}{\partial y}\right)  +\Omega_{Ti}-\Omega_{ni},\tag{A38}\\
W_{e} &  =\gamma\frac{\partial}{\partial t}+\Omega_{Te}-\Omega_{ne}.\nonumber
\end{align}
We remind the reader that the temperatures of the ions and electrons are
considered to be equal one another. We retain different notations for the
control of the symmetry of the ion and electron contribution. Analogously, we
find the following values:%
\begin{align}
G_{4}-DT_{i0}\left(  \frac{\partial}{\partial t}+v_{i0y}\frac{\partial
}{\partial y}\right)  ^{-1} &  =-T_{i0}\left(  W_{i}V_{e}+W_{i}\Omega
_{ei}+V_{e}\Omega_{ie}\right)  \frac{\partial}{\partial t},\tag{A39}\\
G_{1}-DT_{e0}\left(  \frac{\partial}{\partial t}\right)  ^{-1} &
=-T_{e0}\left(  W_{e}V_{i}+W_{e}\Omega_{ie}+V_{i}\Omega_{ei}\right)  \left(
\frac{\partial}{\partial t}+v_{i0y}\frac{\partial}{\partial y}\right)
,\nonumber
\end{align}
where%
\begin{align}
V_{i} &  =\left(  \frac{\partial}{\partial t}+v_{i0y}\frac{\partial}{\partial
y}\right)  +\Omega_{Ti},\tag{A40}\\
V_{e} &  =\frac{\partial}{\partial t}+\Omega_{Te}.\nonumber
\end{align}
\bigskip
\subsection{Operator $L$}
Let us find the operator $L$ given by Equation (A22). Using Equations (A17)
and (A19), we obtain
\begin{align}
L &  =\Omega_{i}^{2}\Omega_{e}^{2}+\frac{1}{m_{i}}\Omega_{e}^{2}\left[
\frac{G_{4}}{D}\left(  \frac{\partial}{\partial t}+v_{i0y}\frac{\partial
}{\partial y}\right)  -T_{i0}\right]  \mathbf{\nabla}^{2}+\frac{1}{m_{e}%
}\Omega_{i}^{2}\left(  \frac{G_{1}}{D}\frac{\partial}{\partial t}%
-T_{e0}\right)  \mathbf{\nabla}^{2}\tag{A41}\\
&  -\frac{1}{m_{i}m_{e}D}\left(  \frac{\partial}{\partial t}+v_{i0y}%
\frac{\partial}{\partial y}\right)  \mathbf{\nabla}^{4}K.\nonumber
\end{align}
The expressions containing in this equation are given by Equations (A37)-(A40).
\bigskip
\subsection{Simplified equations for ion and electron velocities
via\textit{\ }$\mathbf{E}_{1}$}
We now substitute expressions for $A_{1,2i}$ given by Equations (A31) and
(A32) into Equations (A23) and (A24). Then, we replace the values
$\mathbf{F}_{j1}$ and $\Phi_{i,e1}$ by their expressions through
$\mathbf{E}_{1}$ which are given by
\begin{align}
F_{j1x} &  =\frac{q_{j}}{m_{j}}\left[  E_{1x}+v_{j0y}\left(  \frac{\partial
}{\partial t}\right)  ^{-1}\left(  \frac{\partial E_{1x}}{\partial y}%
-\frac{\partial E_{1y}}{\partial x}\right)  \right]  ,\tag{A42}\\
F_{j1y} &  =\frac{q_{j}}{m_{j}}E_{1y}\nonumber
\end{align}
and%
\begin{align}
\Phi_{i1} &  =-\frac{q_{i}}{m_{i}}\left(  \omega_{ci}-v_{i0y}\frac{\partial
}{\partial x}\right)  \left(  \frac{\partial}{\partial t}+v_{i0y}%
\frac{\partial}{\partial y}\right)  \left(  \frac{\partial}{\partial
t}\right)  ^{-1}\left(  \frac{\partial E_{1x}}{\partial y}-\frac{\partial
E_{1y}}{\partial x}\right)  \tag{A43}\\
&  +\frac{q_{i}}{m_{i}}\left(  \frac{\partial}{\partial t}+v_{i0y}%
\frac{\partial}{\partial y}\right)  \mathbf{\nabla\cdot E}_{1},\nonumber\\
\Phi_{e1} &  =-\frac{q_{e}}{m_{e}}\omega_{ce}\left(  \frac{\partial E_{1x}%
}{\partial y}-\frac{\partial E_{1y}}{\partial x}\right)  +\frac{q_{e}}{m_{e}%
}\frac{\partial}{\partial t}\mathbf{\nabla\cdot E}_{1}.\nonumber
\end{align}
For obtaining Equations (A42) and (A43), we have used Equations (A2) and (8).
As a result, we will have the following equations for $v_{i1x}$ and $v_{i1y}$:%
\begin{align}
v_{i1x} &  =-\frac{q_{i}}{m_{i}}\frac{\Omega_{e}^{2}}{\Omega_{i}^{2}}%
\frac{L_{ix}}{L}\lambda_{i}\left[  a_{i}\left(  \frac{\partial E_{1y}%
}{\partial x}-\frac{\partial E_{1x}}{\partial y}\right)  +\left(
\frac{\partial}{\partial t}+v_{i0y}\frac{\partial}{\partial y}\right)
\mathbf{\nabla\cdot E}_{1}\right]  \tag{A44}\\
&  +\frac{q_{e}}{m_{e}}\frac{L_{ix}}{L}\mu_{i}\left[  \omega_{ce}\left(
\frac{\partial E_{1y}}{\partial x}-\frac{\partial E_{1x}}{\partial y}\right)
+\frac{\partial}{\partial t}\mathbf{\nabla\cdot E}_{1}\right]  \nonumber\\
&  +\frac{q_{i}}{m_{i}\Omega_{i}^{2}}\left(  \frac{\partial}{\partial
t}+v_{i0y}\frac{\partial}{\partial y}\right)  ^{2}\left(  \frac{\partial
}{\partial t}\right)  ^{-1}E_{1x}\nonumber\\
&  +\frac{q_{i}}{m_{i}\Omega_{i}^{2}}\left[  \omega_{ci}-v_{i0y}\frac
{\partial}{\partial x}\left(  \frac{\partial}{\partial t}+v_{i0y}%
\frac{\partial}{\partial y}\right)  \left(  \frac{\partial}{\partial
t}\right)  ^{-1}\right]  E_{1y}\nonumber
\end{align}
and%
\begin{align}
v_{i1y} &  =-\frac{q_{i}}{m_{i}}\frac{\Omega_{e}^{2}}{\Omega_{i}^{2}}%
\frac{L_{iy}}{L}\lambda_{i}\left[  a_{i}\left(  \frac{\partial E_{1y}%
}{\partial x}-\frac{\partial E_{1x}}{\partial y}\right)  +\left(
\frac{\partial}{\partial t}+v_{i0y}\frac{\partial}{\partial y}\right)
\mathbf{\nabla\cdot E}_{1}\right]  \tag{A45}\\
&  +\frac{q_{e}}{m_{e}}\frac{L_{iy}}{L}\mu_{i}\left[  \omega_{ce}\left(
\frac{\partial E_{1y}}{\partial x}-\frac{\partial E_{1x}}{\partial y}\right)
+\frac{\partial}{\partial t}\mathbf{\nabla\cdot E}_{1}\right]  \nonumber\\
&  -\frac{q_{i}}{m_{i}}\frac{\omega_{ci}}{\Omega_{i}^{2}}\left(
\frac{\partial}{\partial t}+v_{i0y}\frac{\partial}{\partial y}\right)  \left(
\frac{\partial}{\partial t}\right)  ^{-1}E_{1x}\nonumber\\
&  +\frac{q_{i}}{m_{i}\Omega_{i}^{2}}\left[  \omega_{ci}v_{i0y}\frac{\partial
}{\partial x}\left(  \frac{\partial}{\partial t}\right)  ^{-1}+\left(
\frac{\partial}{\partial t}+v_{i0y}\frac{\partial}{\partial y}\right)
\right]  E_{1y},\nonumber
\end{align}
where notations are
\begin{align}
\lambda_{i} &  =\frac{1}{m_{i}}\left[  T_{i0}\left(  \frac{\partial}{\partial
t}+v_{i0y}\frac{\partial}{\partial y}\right)  ^{-1}-\frac{G_{4}}{D}\right]
+\frac{1}{m_{e}m_{i}D\Omega_{e}^{2}}\mathbf{\nabla}^{2}K,\mu_{i}=\frac{G_{3}%
}{m_{i}D},\tag{A46}\\
a_{i} &  =\left(  \omega_{ci}-v_{i0y}\frac{\partial}{\partial x}\right)
\left(  \frac{\partial}{\partial t}+v_{i0y}\frac{\partial}{\partial y}\right)
\left(  \frac{\partial}{\partial t}\right)  ^{-1}.\nonumber
\end{align}
For the electron velocity, using Equations (A26), (A27), (A34), and (A35), we
obtain%
\begin{align}
v_{e1x} &  =-\frac{q_{e}}{m_{e}}\frac{\Omega_{i}^{2}}{\Omega_{e}^{2}}%
\frac{L_{ex}}{L}\lambda_{e}\left[  \omega_{ce}\left(  \frac{\partial E_{1y}%
}{\partial x}-\frac{\partial E_{1x}}{\partial y}\right)  +\frac{\partial
}{\partial t}\mathbf{\nabla\cdot E}_{1}\right]  \tag{A47}\\
&  +\frac{q_{i}}{m_{i}}\frac{L_{ex}}{L}\mu_{e}\left[  a_{i}\left(
\frac{\partial E_{1y}}{\partial x}-\frac{\partial E_{1x}}{\partial y}\right)
+\left(  \frac{\partial}{\partial t}+v_{i0y}\frac{\partial}{\partial
y}\right)  \mathbf{\nabla\cdot E}_{1}\right]  \nonumber\\
&  +\frac{q_{e}}{m_{e}}\frac{\omega_{ce}}{\Omega_{e}^{2}}E_{1y}+\frac{q_{e}%
}{m_{e}\Omega_{e}^{2}}\frac{\partial E_{1x}}{\partial t}\nonumber
\end{align}
and%
\begin{align}
v_{e1y} &  =-\frac{q_{e}}{m_{e}}\frac{\Omega_{i}^{2}}{\Omega_{e}^{2}}%
\frac{L_{ey}}{L}\lambda_{e}\left[  \omega_{ce}\left(  \frac{\partial E_{1y}%
}{\partial x}-\frac{\partial E_{1x}}{\partial y}\right)  +\frac{\partial
}{\partial t}\mathbf{\nabla\cdot E}_{1}\right]  \tag{A48}\\
&  +\frac{q_{i}}{m_{i}}\frac{L_{ey}}{L}\mu_{e}\left[  a_{i}\left(
\frac{\partial E_{1y}}{\partial x}-\frac{\partial E_{1x}}{\partial y}\right)
+\left(  \frac{\partial}{\partial t}+v_{i0y}\frac{\partial}{\partial
y}\right)  \mathbf{\nabla\cdot E}_{1}\right]  \nonumber\\
&  -\frac{q_{e}}{m_{e}}\frac{\omega_{ce}}{\Omega_{e}^{2}}E_{1x}+\frac{q_{e}%
}{m_{e}\Omega_{e}^{2}}\frac{\partial E_{1y}}{\partial t},\nonumber
\end{align}
where
\begin{align}
\lambda_{e} &  =\frac{1}{m_{e}}\left[  T_{e0}\left(  \frac{\partial}{\partial
t}\right)  ^{-1}-\frac{G_{1}}{D}\right]  +\frac{1}{m_{i}m_{e}\Omega_{i}^{2}%
D}\left(  \frac{\partial}{\partial t}+v_{i0y}\frac{\partial}{\partial
y}\right)  \left(  \frac{\partial}{\partial t}\right)  ^{-1}\mathbf{\nabla
}^{2}K,\tag{A49}\\
\mu_{e} &  =\frac{G_{2}}{m_{e}D}.\nonumber
\end{align}
\bigskip
\subsection{Perturbed plasma currents}
We now make use of obtained ion and electron velocities to find perturbed
plasma currents $j_{pl1x}=q_{i}n_{i0}v_{i1x}+q_{e}n_{e0}v_{e1x}$ and
$j_{pl1y}=q_{i}n_{i0}v_{i1y}+q_{i}n_{i1}v_{i0y}+q_{e}n_{e0}v_{e1y}$ in a
general form. From Equations (A44) and (A47), we will have%
\begin{align}
4\pi\left(  \frac{\partial}{\partial t}\right)  ^{-1}j_{pl1x} &  =\alpha
_{x}\left(  \frac{\partial E_{1y}}{\partial x}-\frac{\partial E_{1x}}{\partial
y}\right)  -\beta_{x}\frac{\partial E_{1y}}{\partial x}+\delta_{x}%
\mathbf{\nabla\cdot E}_{1}\tag{A50}\\
&  +\frac{\omega_{pi}^{2}}{\Omega_{i}^{2}}\left(  \frac{\partial}{\partial
t}+v_{i0y}\frac{\partial}{\partial y}\right)  ^{2}\left(  \frac{\partial
}{\partial t}\right)  ^{-2}E_{1x}+\frac{\omega_{pe}^{2}}{\Omega_{e}^{2}}%
E_{1x}\nonumber\\
&  +\left(  \frac{\omega_{pi}^{2}\omega_{ci}}{\Omega_{i}^{2}}+\frac
{\omega_{pe}^{2}\omega_{ce}}{\Omega_{e}^{2}}\right)  \left(  \frac{\partial
}{\partial t}\right)  ^{-1}E_{1y}.\nonumber
\end{align}
Here,%
\begin{align}
\alpha_{x} &  =\frac{1}{L}\left[  \omega_{pi}^{2}L_{ix}\left(  \frac
{q_{e}m_{i}}{q_{i}m_{e}}\mu_{i}\omega_{ce}-\frac{\Omega_{e}^{2}}{\Omega
_{i}^{2}}\lambda_{i}a_{i}\right)  +\omega_{pe}^{2}L_{ex}\left(  \frac
{q_{i}m_{e}}{q_{e}m_{i}}\mu_{e}a_{i}-\frac{\Omega_{i}^{2}}{\Omega_{e}^{2}%
}\lambda_{e}\omega_{ce}\right)  \right]  \left(  \frac{\partial}{\partial
t}\right)  ^{-1},\tag{A51}\\
\delta_{x} &  =\omega_{pi}^{2}\frac{L_{ix}}{L}\left[  \frac{q_{e}m_{i}}%
{q_{i}m_{e}}\mu_{i}-\frac{\Omega_{e}^{2}}{\Omega_{i}^{2}}\lambda_{i}\left(
\frac{\partial}{\partial t}+v_{i0y}\frac{\partial}{\partial y}\right)  \left(
\frac{\partial}{\partial t}\right)  ^{-1}\right]  \nonumber\\
&  +\omega_{pe}^{2}\frac{L_{ex}}{L}\left[  \frac{q_{i}m_{e}}{q_{e}m_{i}}%
\mu_{e}\left(  \frac{\partial}{\partial t}+v_{i0y}\frac{\partial}{\partial
y}\right)  \left(  \frac{\partial}{\partial t}\right)  ^{-1}-\frac{\Omega
_{i}^{2}}{\Omega_{e}^{2}}\lambda_{e}\right]  ,\nonumber\\
\beta_{x} &  =\frac{\omega_{pi}^{2}}{\Omega_{i}^{2}}v_{i0y}\left(
\frac{\partial}{\partial t}+v_{i0y}\frac{\partial}{\partial y}\right)  \left(
\frac{\partial}{\partial t}\right)  ^{-2},\nonumber
\end{align}
and $\omega_{pj}=\left(  4\pi n_{j0}q_{j}^{2}/m_{j}\right)  ^{1/2}$ is the
plasma frequency. The values $\lambda_{i,e}$, $\mu_{i,e}$, and $a_{i}$ are
given by Equations (A46) and (A49). Using Equations (2), (A44), (A45), and
(A48), we further find
\begin{align}
4\pi\left(  \frac{\partial}{\partial t}\right)  ^{-1}j_{pl1y} &  =\left(
\alpha_{y}+\eta_{1}\right)  \left(  \frac{\partial E_{1y}}{\partial x}%
-\frac{\partial E_{1x}}{\partial y}\right)  -\beta_{x}\frac{\partial E_{1x}%
}{\partial x}+\beta_{y}\frac{\partial E_{1y}}{\partial x}+\left(  \delta
_{y}+\eta_{2}\right)  \mathbf{\nabla\cdot E}_{1}\tag{A52}\\
&  -\left(  \frac{\omega_{pi}^{2}\omega_{ci}}{\Omega_{i}^{2}}+\frac
{\omega_{pe}^{2}\omega_{ce}}{\Omega_{e}^{2}}\right)  \left(  \frac{\partial
}{\partial t}\right)  ^{-1}E_{1x}+\left(  \frac{\omega_{pi}^{2}}{\Omega
_{i}^{2}}+\frac{\omega_{pe}^{2}}{\Omega_{e}^{2}}\right)  E_{1y},\nonumber
\end{align}
where%
\begin{align}
\alpha_{y} &  =\omega_{pi}^{2}\frac{L_{iy}}{L}\left(  \frac{q_{e}m_{i}}%
{q_{i}m_{e}}\mu_{i}\omega_{ce}-\frac{\Omega_{e}^{2}}{\Omega_{i}^{2}}%
\lambda_{i}a_{i}\right)  \left(  \frac{\partial}{\partial t}+v_{i0y}%
\frac{\partial}{\partial y}\right)  ^{-1}\tag{A53}\\
&  +\omega_{pe}^{2}\frac{L_{ey}}{L}\left(  \frac{q_{i}m_{e}}{q_{e}m_{i}}%
\mu_{e}a_{i}-\frac{\Omega_{i}^{2}}{\Omega_{e}^{2}}\lambda_{e}\omega
_{ce}\right)  \left(  \frac{\partial}{\partial t}\right)  ^{-1},\nonumber\\
\delta_{y} &  =\omega_{pi}^{2}\frac{L_{iy}}{L}\left[  \frac{q_{e}m_{i}}%
{q_{i}m_{e}}\mu_{i}\left(  \frac{\partial}{\partial t}+v_{i0y}\frac{\partial
}{\partial y}\right)  ^{-1}\frac{\partial}{\partial t}-\frac{\Omega_{e}^{2}%
}{\Omega_{i}^{2}}\lambda_{i}\right]  \nonumber\\
&  +\omega_{pe}^{2}\frac{L_{ey}}{L}\left[  \frac{q_{i}m_{e}}{q_{e}m_{i}}%
\mu_{e}\left(  \frac{\partial}{\partial t}+v_{i0y}\frac{\partial}{\partial
y}\right)  \left(  \frac{\partial}{\partial t}\right)  ^{-1}-\frac{\Omega
_{i}^{2}}{\Omega_{e}^{2}}\lambda_{e}\right]  ,\nonumber\\
\eta_{1} &  =\omega_{pi}^{2}v_{i0y}\frac{L_{ix}}{L}\left(  \frac{\Omega
_{e}^{2}}{\Omega_{i}^{2}}\lambda_{i}a_{i}-\frac{q_{e}m_{i}}{q_{i}m_{e}}\mu
_{i}\omega_{ce}\right)  \left(  \frac{\partial}{\partial t}+v_{i0y}%
\frac{\partial}{\partial y}\right)  ^{-1}\left(  \frac{\partial}{\partial
t}\right)  ^{-1}\frac{\partial}{\partial x},\nonumber\\
\eta_{2} &  =\omega_{pi}^{2}v_{i0y}\frac{L_{ix}}{L}\left[  \frac{\Omega
_{e}^{2}}{\Omega_{i}^{2}}\lambda_{i}\left(  \frac{\partial}{\partial
t}\right)  ^{-1}-\frac{q_{e}m_{i}}{q_{i}m_{e}}\mu_{i}\left(  \frac{\partial
}{\partial t}+v_{i0y}\frac{\partial}{\partial y}\right)  ^{-1}\right]
\frac{\partial}{\partial x},\nonumber\\
\beta_{y} &  =\frac{\omega_{pi}^{2}}{\Omega_{i}^{2}}v_{i0y}^{2}\left(
\frac{\partial}{\partial t}\right)  ^{-2}\frac{\partial}{\partial x}.\nonumber
\end{align}
We can rewrite Equations (A50) and (A52) in the form%
\begin{equation}
4\pi\left(  \frac{\partial}{\partial t}\right)  ^{-1}j_{pl1x}=\varepsilon
_{plxx}E_{1x}+\varepsilon_{plxy}E_{1y}\tag{A54}%
\end{equation}
and%
\begin{equation}
4\pi\left(  \frac{\partial}{\partial t}\right)  ^{-1}j_{pl1y}=\varepsilon
_{plyx}E_{1x}+\varepsilon_{plyy}E_{1y},\tag{A55}%
\end{equation}
where the components of the plasma dielectric permeability tensor are given by%
\begin{align}
\varepsilon_{plxx} &  =-\alpha_{x}\frac{\partial}{\partial y}+\delta_{x}%
\frac{\partial\ }{\partial x}+\frac{\omega_{pi}^{2}}{\Omega_{i}^{2}}\left(
\frac{\partial}{\partial t}+v_{i0y}\frac{\partial}{\partial y}\right)
^{2}\left(  \frac{\partial}{\partial t}\right)  ^{-2}+\frac{\omega_{pe}^{2}%
}{\Omega_{e}^{2}},\tag{A56}\\
\varepsilon_{plxy} &  =\alpha_{x}\frac{\partial}{\partial x}-\beta_{x}%
\frac{\partial}{\partial x}+\delta_{x}\frac{\partial}{\partial y}+\left(
\frac{\omega_{pi}^{2}\omega_{ci}}{\Omega_{i}^{2}}+\frac{\omega_{pe}^{2}%
\omega_{ce}}{\Omega_{e}^{2}}\right)  \left(  \frac{\partial}{\partial
t}\right)  ^{-1},\nonumber\\
\varepsilon_{plyx} &  =-\left(  \alpha_{y}+\eta_{1}\right)  \frac{\partial
}{\partial y}-\beta_{x}\frac{\partial}{\partial x}+\left(  \delta_{y}+\eta
_{2}\right)  \frac{\partial}{\partial x}-\left(  \frac{\omega_{pi}^{2}%
\omega_{ci}}{\Omega_{i}^{2}}+\frac{\omega_{pe}^{2}\omega_{ce}}{\Omega_{e}^{2}%
}\right)  \left(  \frac{\partial}{\partial t}\right)  ^{-1},\nonumber\\
\varepsilon_{plyy} &  =\left(  \alpha_{y}+\eta_{1}\right)  \frac{\partial
}{\partial x}+\beta_{y}\frac{\partial}{\partial x}+\left(  \delta_{y}+\eta
_{2}\right)  \frac{\partial}{\partial y}+\frac{\omega_{pi}^{2}}{\Omega_{i}%
^{2}}+\frac{\omega_{pe}^{2}}{\Omega_{e}^{2}}.\nonumber
\end{align}
Using Equations (A51) and (A53), we can find $\varepsilon_{plij}$ in specific cases.
\bigskip
\section{Appendix}
\bigskip
\subsection{Perturbed velocity of cosmic rays}
The linearized Equation (5) for the cold, nonrelativistic, $T_{cr}\ll
m_{cr}c^{2}$, cosmic rays takes the form%
\begin{equation}
\gamma_{cr0}\left(  \frac{\partial}{\partial t}+u_{cr}\frac{\partial}{\partial
y}\right)  \left(  \mathbf{v}_{cr1}+\gamma_{cr0}^{2}\frac{\mathbf{u}%
_{cr}u_{cr}}{c^{2}}v_{cr1y}\right)  =-\frac{\mathbf{\nabla}p_{cr1}}%
{m_{cr}n_{cr0}}+\mathbf{F}_{cr1}+\frac{q_{cr}}{m_{cr}c}\mathbf{v}_{cr1}%
\times\mathbf{B}_{0},\tag{B1}%
\end{equation}
where%
\begin{equation}
\mathbf{F}_{cr1}=\frac{q_{cr}}{m_{cr}}\left(  \mathbf{E}_{1}\mathbf{+}\frac
{1}{c}\mathbf{u}_{cr}\times\mathbf{B}_{1}\right)  .\tag{B2}%
\end{equation}
For obtaining Equation (B1), we have used that $\mathbf{u}_{cr}$ is directed
along the $y$-axis and $\gamma_{cr1}=\gamma_{cr0}^{3}u_{cr}v_{cr1y}/c^{2}$,
where $\gamma_{cr0}=\left(  1-u_{cr}^{2}/c^{2}\right)  ^{-1/2}$. From Equation
(B1), we find the following equations for $v_{cr1x,y}$:%
\begin{equation}
\gamma_{cr0}\left(  \frac{\partial}{\partial t}+u_{cr}\frac{\partial}{\partial
y}\right)  v_{cr1x}=-\frac{1}{m_{cr}n_{cr0}}\frac{\partial p_{cr1}}{\partial
x}+F_{cr1x}+\omega_{ccr}v_{cr1y}\ \tag{B3}%
\end{equation}
and
\begin{equation}
\gamma_{cr0}^{3}\left(  \frac{\partial}{\partial t}+u_{cr}\frac{\partial
}{\partial y}\right)  v_{cr1y}=-\frac{1}{m_{cr}n_{cr0}}\frac{\partial p_{cr1}%
}{\partial y}+F_{cr1y}-\omega_{ccr}v_{cr1x},\tag{B4}%
\end{equation}
where $\omega_{ccr}=q_{cr}B_{0}/m_{cr}c$ is the cyclotron frequency of the
cosmic-ray particles. Solutions of Equations (B3) and (B4) have the form%
\begin{equation}
\Omega_{cr}^{2}v_{cr1x}=\frac{1}{m_{cr}n_{cr0}}L_{1crx}p_{cr1}+\omega
_{ccr}\ F_{cr1y}+\gamma_{cr0}^{3}\left(  \frac{\partial}{\partial t}%
+u_{cr}\frac{\partial}{\partial y}\right)  F_{cr1x}\tag{B5}%
\end{equation}
and%
\begin{equation}
\Omega_{cr}^{2}v_{cr1y}=\frac{1}{m_{cr}n_{cr0}}L_{1cry}p_{cr1}-\omega
_{ccr}F_{cr1x}+\gamma_{cr0}\left(  \frac{\partial}{\partial t}+u_{cr}%
\frac{\partial}{\partial y}\right)  F_{cr1y},\tag{B6}%
\end{equation}
where
\begin{align}
\Omega_{cr}^{2} &  =\gamma_{cr0}^{4}\left(  \frac{\partial}{\partial t}%
+u_{cr}\frac{\partial}{\partial y}\right)  ^{2}+\omega_{ccr}^{2},\tag{B7}\\
L_{1crx} &  =-\omega_{ccr}\ \frac{\partial}{\partial y}-\gamma_{cr0}%
^{3}\left(  \frac{\partial}{\partial t}+u_{cr}\frac{\partial}{\partial
y}\right)  \frac{\partial}{\partial x},\nonumber\\
L_{1cry} &  =\omega_{ccr}\frac{\partial}{\partial x}-\gamma_{cr0}\left(
\frac{\partial}{\partial t}+u_{cr}\frac{\partial}{\partial y}\right)
\frac{\partial}{\partial y}.\nonumber
\end{align}
\bigskip
\subsection{Equation for perturbed cosmic-ray pressure}
From Equation (6) in the linear approximation, we obtain the perturbed
cosmic-ray pressure%
\begin{equation}
p_{cr1}=p_{cr0}\Gamma_{cr}\left(  \frac{n_{cr1}}{n_{cr0}}-\frac{\gamma_{cr1}%
}{\gamma_{cr0}}\right)  .\tag{B8}%
\end{equation}
Using the linearized continuity equation (2) for cosmic rays and expression
for $\gamma_{cr1}$, we find that $p_{cr1}$ is given by
\begin{equation}
p_{cr1}=-p_{cr0}\Gamma_{cr}\left[  \left(  \frac{\partial}{\partial t}%
+u_{cr}\frac{\partial}{\partial y}\right)  ^{-1}\mathbf{\nabla\cdot v}%
_{cr1}+\gamma_{cr0}^{2}\frac{u_{cr}}{c^{2}}v_{cr1y}\right]  .\tag{B9}%
\end{equation}
From Equations (B5) and (B6), we obtain the expression for $\mathbf{\nabla
\cdot v}_{cr1}$ which is substituted together with the velocity $v_{cr1y}$
into Equation (B9). As a result, we have%
\begin{equation}
L_{2cr}p_{cr1}=-p_{cr0}\Gamma_{cr}\Phi_{cr1}.\tag{B10}%
\end{equation}
Here,%
\begin{align}
L_{2cr} &  =\Omega_{cr}^{2}-\gamma_{cr0}c_{scr}^{2}L_{1cr}+\gamma_{cr0}%
^{2}\frac{u_{cr}}{c^{2}}c_{scr}^{2}L_{1cry},\tag{B11}\\
\Phi_{cr1} &  =-L_{3crx}F_{cr1x}+L_{3cry}F_{cr1y},\nonumber
\end{align}
where $c_{scr}=\left(  p_{cr0}\Gamma_{cr}/m_{cr}n_{cr0}\right)  ^{1/2}$ is the
cosmic-ray sound speed defined by the rest mass and%
\begin{align}
L_{1cr} &  =\gamma_{cr0}^{2}\frac{\partial^{2}}{\partial x^{2}}+\frac
{\partial^{2}}{\partial y^{2}},\tag{B12}\\
L_{3crx} &  =\omega_{ccr}\gamma_{cr0}^{2}\left(  \frac{u_{cr}}{c^{2}}%
\frac{\partial}{\partial t}+\frac{\partial}{\partial y}\right)  \left(
\frac{\partial}{\partial t}+u_{cr}\frac{\partial}{\partial y}\right)
^{-1}-\gamma_{cr0}^{3}\frac{\partial}{\partial x},\nonumber\\
L_{3cry} &  =\omega_{ccr}\left(  \frac{\partial}{\partial t}+u_{cr}%
\frac{\partial}{\partial y}\right)  ^{-1}\ \frac{\partial}{\partial x}%
+\gamma_{cr0}^{3}\left(  \frac{u_{cr}}{c^{2}}\frac{\partial}{\partial t}%
+\frac{\partial}{\partial y}\right)  .\nonumber
\end{align}
\bigskip
\subsection{Equations for cosmic ray velocities via\textit{\ }%
$\mathbf{F}_{cr1}$}
Substituting Equations (B10) and (B11) into Equations (B5) and (B6), we find%
\begin{equation}
\Omega_{cr}^{2}v_{cr1x}=\left[  c_{scr}^{2}\frac{L_{1crx}}{L_{2cr}}%
L_{3crx}+\gamma_{cr0}^{3}\left(  \frac{\partial}{\partial t}+u_{cr}%
\frac{\partial}{\partial y}\right)  \right]  F_{cr1x}+\left(  -c_{scr}%
^{2}\frac{L_{1crx}}{L_{2cr}}L_{3cry}+\omega_{ccr}\right)  \ F_{cr1y}\tag{B13}%
\end{equation}
and%
\begin{equation}
\Omega_{cr}^{2}v_{cr1y}=\left(  c_{scr}^{2}\frac{L_{1cry}}{L_{2cr}}%
L_{3crx}-\omega_{ccr}\right)  F_{cr1x}+\left[  -c_{scr}^{2}\frac{L_{1cry}%
}{L_{2cr}}L_{3cry}+\gamma_{cr0}\left(  \frac{\partial}{\partial t}+u_{cr}%
\frac{\partial}{\partial y}\right)  \right]  F_{cr1y}.\tag{B14}%
\end{equation}
\bigskip
\subsection{Equations for cosmic-ray velocities via\textit{\ }%
$\mathbf{E}_{1}$}
From Equation (B2), we have%
\begin{align}
F_{cr1x} &  =\frac{q_{cr}}{m_{cr}}\left[  E_{1x}+u_{cr}\left(  \frac{\partial
}{\partial t}\right)  ^{-1}\left(  \frac{\partial E_{1x}}{\partial y}%
-\frac{\partial E_{1y}}{\partial x}\right)  \right]  ,\tag{B15}\\
F_{cr1y} &  =\frac{q_{cr}}{m_{cr}}E_{1y}.\nonumber
\end{align}
Substituting Equation (B15) into Equations (B13) and (B14), we obtain%
\begin{align}
v_{cr1x} &  =\frac{q_{cr}}{m_{cr}\Omega_{cr}^{2}}\left[  a_{crx}+\gamma
_{cr0}^{3}\left(  \frac{\partial}{\partial t}+u_{cr}\frac{\partial}{\partial
y}\right)  ^{2}\left(  \frac{\partial}{\partial t}\right)  ^{-1}\right]
E_{1x}\tag{B16}\\
&  +\frac{q_{cr}}{m_{cr}\Omega_{cr}^{2}}\left[  -b_{crx}+\omega_{ccr}%
-\gamma_{cr0}^{3}u_{cr}\frac{\partial}{\partial x}\left(  \frac{\partial
}{\partial t}+u_{cr}\frac{\partial}{\partial y}\right)  \left(  \frac
{\partial}{\partial t}\right)  ^{-1}\right]  E_{1y}\nonumber
\end{align}
and%
\begin{align}
v_{cr1y} &  =\frac{q_{cr}}{m_{cr}\Omega_{cr}^{2}}\left[  a_{cry}-\omega
_{ccr}\left(  \frac{\partial}{\partial t}+u_{cr}\frac{\partial}{\partial
y}\right)  \left(  \frac{\partial}{\partial t}\right)  ^{-1}\right]
E_{1x}\tag{B17}\\
&  +\frac{q_{cr}}{m_{cr}\Omega_{cr}^{2}}\left[  -b_{cry}+\omega_{ccr}%
u_{cr}\frac{\partial}{\partial x}\left(  \frac{\partial}{\partial t}\right)
^{-1}+\gamma_{cr0}\left(  \frac{\partial}{\partial t}+u_{cr}\frac{\partial
}{\partial y}\right)  \right]  E_{1y},\nonumber
\end{align}
where%
\begin{align}
a_{crx} &  =c_{scr}^{2}\frac{L_{1crx}}{L_{2cr}}L_{3crx}\left(  \frac{\partial
}{\partial t}+u_{cr}\frac{\partial}{\partial y}\right)  \left(  \frac
{\partial}{\partial t}\right)  ^{-1},\tag{B18}\\
b_{crx} &  =c_{scr}^{2}\frac{L_{1crx}}{L_{2cr}}\left[  L_{3cry}+L_{3crx}%
u_{cr}\frac{\partial}{\partial x}\left(  \frac{\partial}{\partial t}\right)
^{-1}\right]  ,\nonumber\\
a_{cry} &  =c_{scr}^{2}\frac{L_{1cry}}{L_{2cr}}L_{3crx}\left(  \frac{\partial
}{\partial t}+u_{cr}\frac{\partial}{\partial y}\right)  \left(  \frac
{\partial}{\partial t}\right)  ^{-1},\nonumber\\
b_{cry} &  =c_{scr}^{2}\frac{L_{1cry}}{L_{2cr}}\left[  L_{3cry}+L_{3crx}%
u_{cr}\frac{\partial}{\partial x}\left(  \frac{\partial}{\partial t}\right)
^{-1}\right]  .\nonumber
\end{align}
The operators $L_{1crx,y}$, $L_{2cr}$, and $L_{3crx,y}$ containing in Equation
(B18) are given by Equations (B7), (B11), and (B12), respectively.
\bigskip
\subsection{Perturbed cosmic ray current}
We now find the components of the perturbed cosmic ray current $j_{cr1x}%
=q_{cr}n_{cr0}v_{cr1x}$ and $j_{cr1y}=$ $q_{cr}n_{cr0}v_{cr1y}+q_{cr}%
n_{cr1}u_{cr}$. Using Equations (B16) and (B17) and the continuity equation
(2) in the linear approximation, we find%
\begin{equation}
4\pi\left(  \frac{\partial}{\partial t}\right)  ^{-1}j_{cr1x}=\varepsilon
_{crxx}E_{1x}+\varepsilon_{crxy}E_{1y}\tag{B19}%
\end{equation}
and
\begin{equation}
4\pi\left(  \frac{\partial}{\partial t}\right)  ^{-1}j_{cr1y}=\varepsilon
_{cryx}E_{1x}+\varepsilon_{cryy}E_{1y}.\tag{B20}%
\end{equation}
The components of the dielectric permeability tensor are the following:%
\begin{align}
\varepsilon_{crxx} &  =\frac{\omega_{pcr}^{2}}{\Omega_{cr}^{2}}\left[
a_{crx}+\gamma_{cr0}^{3}\left(  \frac{\partial}{\partial t}+u_{cr}%
\frac{\partial}{\partial y}\right)  ^{2}\left(  \frac{\partial}{\partial
t}\right)  ^{-1}\right]  \left(  \frac{\partial}{\partial t}\right)
^{-1},\tag{B21}\\
\varepsilon_{crxy} &  =\frac{\omega_{pcr}^{2}}{\Omega_{cr}^{2}}\left[
-b_{crx}+\omega_{ccr}-\gamma_{cr0}^{3}u_{cr}\frac{\partial}{\partial x}\left(
\frac{\partial}{\partial t}+u_{cr}\frac{\partial}{\partial y}\right)  \left(
\frac{\partial}{\partial t}\right)  ^{-1}\right]  \left(  \frac{\partial
}{\partial t}\right)  ^{-1},\nonumber\\
\varepsilon_{cryx} &  =\frac{\omega_{pcr}^{2}}{\Omega_{cr}^{2}}\left[  \left(
a_{cry}\frac{\partial}{\partial t}-a_{crx}u_{cr}\frac{\partial}{\partial
x}\right)  \left(  \frac{\partial}{\partial t}+u_{cr}\frac{\partial}{\partial
y}\right)  ^{-1}\right]  \left(  \frac{\partial}{\partial t}\right)
^{-1}\nonumber\\
&  -\frac{\omega_{pcr}^{2}}{\Omega_{cr}^{2}}\left[  \omega_{ccr}+\gamma
_{cr0}^{3}u_{cr}\frac{\partial}{\partial x}\left(  \frac{\partial}{\partial
t}+u_{cr}\frac{\partial}{\partial y}\right)  \left(  \frac{\partial}{\partial
t}\right)  ^{-1}\right]  \left(  \frac{\partial}{\partial t}\right)
^{-1},\nonumber\\
\varepsilon_{cryy} &  =\frac{\omega_{pcr}^{2}}{\Omega_{cr}^{2}}\left[  \left(
-b_{cry}\frac{\partial}{\partial t}+b_{crx}u_{cr}\frac{\partial}{\partial
x}\right)  \left(  \frac{\partial}{\partial t}+u_{cr}\frac{\partial}{\partial
y}\right)  ^{-1}\right]  \left(  \frac{\partial}{\partial t}\right)
^{-1}\nonumber\\
&  +\frac{\omega_{pcr}^{2}}{\Omega_{cr}^{2}}\gamma_{cr0}\left[  1+\gamma
_{cr0}^{2}u_{cr}^{2}\frac{\partial^{2}}{\partial x^{2}}\left(  \frac{\partial
}{\partial t}\right)  ^{-2}\right]  .\nonumber
\end{align}
\bigskip
\bigskip
\end{appendix}

\end{document}